\documentclass[12pt]{article}
\pdfoutput=1
\usepackage[DIV13]{typearea}
\usepackage{amsmath}
\usepackage{amssymb}
\usepackage{graphicx}
\usepackage{cite}
\usepackage{bbold}
\newcommand{\be}{\beta}

\def\be{\begin{equation}}
\def\ee{\end{equation}}
\def\beq{\begin{equation}}
\def\eeq{\end{equation}}
\def\bc{\begin{center}}
\def\ec{\end{center}}
\def\bea{\begin{eqnarray}}
\def\eea{\end{eqnarray}}

\begin{document}
\begin{titlepage}
\vspace*{-1cm}
\phantom{hep-ph/***} 
\flushright
\hfil{DFPD-2012/TH/21}
\hfil{TUM-HEP-868/12}\\

\vskip 1.5cm
\begin{center}
\mathversion{bold}
{\LARGE\bf Lepton Mixing Parameters from Discrete and CP Symmetries}\\[3mm]
\mathversion{normal}
\vskip .3cm
\end{center}
\vskip 0.5  cm
\begin{center}
{\large Ferruccio Feruglio}~$^{a),b)}$,
{\large Claudia Hagedorn}~$^{a),b),c)}$\\[2mm]
{\large and Robert Ziegler}~$^{d),e)}$
\\
\vskip .7cm
{\footnotesize
$^{a)}$~Dipartimento di Fisica e Astronomia `G.~Galilei', Universit\`a di Padova,\\ Via Marzolo~8, I-35131 Padua, Italy
\vskip .1cm
$^{b)}$~INFN, Sezione di Padova, Via Marzolo~8, I-35131 Padua, Italy
\vskip .1cm
$^{c)}$~SISSA, Via Bonomea 265, I-34136 Trieste, Italy
\vskip .1cm
$^{d)}$~Physik-Department, Technische Universit\"at M\"unchen, 
\\
James-Franck-Strasse, D-85748 Garching, Germany
\\
\vskip .1cm
$^{e)}$~
TUM Institute for Advanced Study, Technische Universit\"at M\"unchen, \\
Lichtenbergstrasse 2a, D-85748 Garching, Germany
\vskip .5cm
\begin{minipage}[l]{.9\textwidth}
\begin{center} 
\textit{E-mail:} 
\tt{feruglio@pd.infn.it}, \tt{hagedorn@pd.infn.it}, \tt{robert.ziegler@ph.tum.de}
\end{center}
\end{minipage}
}
\end{center}
\vskip 1cm
\begin{abstract}
We consider a scenario with three Majorana neutrinos in which a discrete, finite flavour group $G_f$ is combined with a generalized $CP$ transformation. We derive
conditions for consistently defining such a setup. 
We show that in general lepton mixing angles
and CP phases (Dirac as well as Majorana) only depend on one single parameter $\theta$ which can take values between $0$ and $\pi$, if the residual symmetries
are $G_e \subset G_f$ in the charged lepton and $G_\nu=Z_2 \times CP$ in the neutrino sector. 
We perform a comprehensive study for $G_f=S_4$ and find five cases which are phenomenologically interesting. They naturally
lead to  a non-zero reactor mixing angle and all mixing parameters are strongly correlated. Some of the patterns predict maximal atmospheric mixing and
maximal Dirac phase, while others predict trivial Dirac and Majorana phases. 
\end{abstract}
\end{titlepage}
\setcounter{footnote}{0}

\section{Introduction}
\label{intro}

Flavour groups $G_f$ and their peculiar breaking to residual symmetries $G_e$ and $G_\nu$ in the charged lepton and in the neutrino sector, respectively, 
have been applied in the past in order to predict lepton mixing angles and the Dirac phase in a model-independent way, for reviews see \cite{reviews}. One of the most prominent
examples is tri-bimaximal (TB) mixing \cite{TB} which can be derived with the help of the flavour groups $A_4$ \cite{A4TB} and $S_4$ \cite{S4TB}. Recent measurements
of the reactor mixing angle $\theta_{13}$, $\sin\theta_{13} \approx 0.15$, \cite{theta13_exp,global_latest}, however, clearly show that many patterns which have
been discussed in the literature are strongly disfavoured, because they predict vanishing or very small $\theta_{13}$. For this reason, proposals have been made in which 
either the groups $G_f$ are chosen to be large, e.g. $\Delta(96)$ and $\Delta(384)$ \cite{Delta96_384}, or the symmetry is reduced, e.g. $G_\nu=Z_2$ is considered instead of $G_\nu=Z_2 \times Z_2$ 
in the neutrino sector \cite{GnuZ2_acc}.

We explore a different approach in this paper and consider a scenario with three Majorana neutrinos in which a discrete, finite flavour group $G_f$ and a $CP$ symmetry are combined and are broken in such a way that 
the residual symmetry in the neutrino sector is $G_\nu=Z_2 \times CP$ with $Z_2$ being a subgroup of $G_f$. The residual symmetry $G_e \subset G_f$ in the charged lepton sector is 
- as in preceding approaches \cite{S4TB,Delta96_384,GnuZ2_acc} - chosen as cyclic symmetry (or product thereof) which allows to distinguish between the three generations
of charged leptons. We show for a general $G_f$ that such a breaking pattern allows to predict lepton mixing angles and Dirac as well as Majorana phases in terms of a 
single real parameter $\theta$. A non-vanishing reactor mixing angle can be easily accommodated and at the same time relations between mixing angles and
CP phases are obtained. In contrast to the scenario without a $CP$ symmetry, we also predict Majorana phases with our present approach. 

We discuss in detail the conditions which have to be fulfilled in order to consistently formulate a setup with a flavour symmetry $G_f$ and a $CP$ symmetry and in order to define the group $G_\nu$
as a direct product of $Z_2$ and $CP$. We also show that in general several independent $CP$ transformations might be compatible with a flavour group $G_f$ which lead, in general, to
physically different results. 
We exemplify our formalism with a comprehensive study 
of $G_f=S_4$ and show several interesting mixing patterns whose predictions for the mixing angles are close to the best fit results \cite{global_latest} for certain values
of the parameter $\theta$. 

The idea to combine a flavour symmetry $G_f$ with a $CP$ symmetry is not new and has been already discussed in some cases in the literature \cite{mutaureflection_HS,mutaureflection_GL,invariant_matrix_elements,simplest_nu_mass_S4}.
An interesting example is the so-called $\mu\tau$ reflection symmetry which is a combination of the canonical $CP$ transformation and the $\mu\tau$ exchange symmetry.
This generalized $CP$ transformation permutes a muon neutrino (antineutrino) and a tau antineutrino (neutrino). 
If imposed on the neutrino mass matrix in the basis in which the charged lepton mass matrix is diagonal, it forces the elements of the second and third rows of the Pontecorvo-Maki-Nakagawa-Sakata
(PMNS) mixing matrix to have the same absolute values. As a consequence, one finds
$\sin\theta_{23}=\cos\theta_{23}$ and $\sin\theta_{13} \sin 2 \theta_{12} \cos\delta=0$. Thus, maximal atmospheric mixing is predicted and, in view of the
latest global fits \cite{global_latest}, also the Dirac phase $\delta$ has to be maximal. Recently, also the combination of the flavour group $S_4$ with a certain $CP$ transformation has been discussed in two models \cite{S4andCP_MN,simplest_nu_mass_S4}.

Other contexts in which flavour symmetries and CP violation appear together are: the idea of so-called geometrical CP violation in which the potential of certain scalars is constrained by a
flavour group in such a way that their vacuum expectation values spontaneously break $CP$ symmetry with phases independent of the parameters of the Lagrangian \cite{geometrical_CPV}; the accidental presence
of $CP$ symmetries has been noticed in potentials invariant under (single- and double-valued) dihedral groups \cite{CP_acc_symm_potential}, while attempts to relate the prediction of CP violation
to particular properties of certain representations of the group $T'$ can be found in \cite{CPV_Tprime}.

The paper is organized as follows: in section \ref{sec2} we first recall several well-known facts about $CP$ transformations and discuss their combination with a flavour group $G_f$
and which conditions have to be fulfilled in order to consistently define such a setup. 
Assuming three generations of Majorana neutrinos, we then present the results for mixing angles and CP phases which are obtained for a general $G_f$ combined with a $CP$ transformation and
 broken to $G_e$ and to $G_\nu=Z_2 \times CP$.
We furthermore study the possibility of and conditions for the presence of an accidental $CP$ symmetry. 
In section \ref{sec3} we present the case $G_f=S_4$  and the different possible $CP$ transformations compatible with all requirements. 
We show that there are only a few independent - and phenomenologically interesting - cases for which we discuss the results for CP phases and mixing angles in detail. 
Apart from that we analyze particular values of the parameter $\theta$ for which the symmetry in the neutrino
sector is enhanced, $G_\nu=Z_2\times Z_2 \times CP$, with $Z_2\times Z_2$ being a subgroup of a finite flavour group containing $S_4$. 
We comment on results which can also be obtained for $G_f=A_4$. We conclude in section \ref{concl}. In appendix \ref{appA} we fix our notations and conventions for the mixing parameters and
show that the invariance of the charged lepton and neutrino mass matrices under a common generalized $CP$ transformation $X$ leads to trivial Dirac and Majorana phases. In appendix \ref{appB} we show that
the mathematical structure of the group arising from the combination of a flavour symmetry $G_f$ and a $CP$ symmetry is in general a semi-direct product of the form $G_{CP}=G_f \rtimes H_{CP}$
with $H_{CP}$ being the group associated with $CP$.

\section{Framework}
\label{sec2}

In this section we recall some
basic properties of $CP$ transformations and of $CP$ invariance in the lepton sector. We then discuss how to combine $CP$ and internal symmetries
and we explain how to use $CP$ and flavour symmetries to constrain lepton mixing parameters. Finally, we comment on the possibility of accidental $CP$ symmetries. 

\subsection{Generalized $CP$ transformations}
\label{CPgen}

We can define a $CP$ transformation on a set of fields, collectively denoted by $\varphi$, as
\be
\varphi'(x)= X \varphi^*(x_{CP})~~~~
\label{X}
\ee
in matrix notation and with $x_{CP}=(x^0,-\vec{x})$. We choose the transformation $X$ to be a constant unitary symmetric matrix, 
\be 
\label{Xcond1}
X X^\dagger = X X^* = \mathbb{1}~~~~.
\ee
In this way,  $CP^2=1$, since eq.(\ref{X}) implies that $(\varphi^*(x_{CP}))'= X^* \varphi (x)$. In theories with internal symmetries, such as a flavour symmetry $G_f$, it is possible
to generalize this requirement on $X$ to $X X^*$ being a transformation belonging to the group of internal symmetries.
We refer to the transformation in eq. (\ref{X}) as a generalized $CP$ transformation \cite{general_CPV}.
For spinors we use a two-component notation and omit the obvious action of $CP$ on the spinor indices (which contains suitable phases such that $CP^2 =1$). 
In the following we also omit the dependence of the fields on the space-time point $x$. 

Let us consider a generalized $CP$ transformation acting on the three generations of lepton doublets $l$
 \be
 l'=X l^*~~~
 \ee
 which fulfills eq.(\ref{Xcond1}).
 In the interaction basis gauge interactions are $CP$ conserving, while the requirement of $CP$ invariance
of the Yukawa interactions, including terms giving rise to neutrino masses, constrains both the charged lepton mass matrix $m_l$\footnote{We use a basis in 
which right-handed (left-handed) fields are on the left-hand (right-hand) side of the charged lepton mass matrix $m_l$.}
 and the neutrino mass matrix $m_\nu$:\footnote{Note that 
 our constraint on the neutrino mass matrix differs by a sign from that in \cite{Branco_review}. This is due to a different definition of the action of $CP$ on the spinor indices.\\
 If $X$ is not symmetric, the conditions in eq.(\ref{invcond}) have to be changed into: $X^\dagger m_l^\dagger m_l X = (m_l^\dagger m_l )^*$ and $X^T m_\nu X = m_\nu^*$.}
\be
\label{invcond}
X^* m_l^\dagger m_l X = (m_l^\dagger m_l )^*~~~, \qquad  \qquad
X m_\nu X = m_\nu^*~~~.
\ee
As we show explicitly in appendix \ref{appA} and is discussed in the literature, see e.g. \cite{Branco_review}, the existence of a $CP$ transformation fulfilling eq.(\ref{invcond}) 
implies that the CP invariants $J_{CP}$ and $I_{1,2}$  (see appendix \ref{appA} for notation and conventions of the mixing parameters)
vanish and thus the Dirac phase $\delta$ as well as the Majorana phases  $\alpha$ and $\beta$ are trivial, i.e.
\be
\label{phasestriv}
\sin\delta=0~~~,~~~\sin \alpha=0~~~,~~~\sin\beta=0~~~ \, .
\ee
If neutrinos are Dirac particles, the second equality in eq.(\ref{invcond}) has to be changed into
\be
\label{Diracnu}
X^* m_\nu^\dagger m_\nu X = (m_\nu^\dagger m_\nu )^* \; .
\ee
Again, as we show in appendix \ref{appA}, the Dirac phase is trivial. This result can also be applied to the quark sector, i.e. the up quark and down quark mass matrices instead of Dirac neutrino and charged lepton mass matrices.

\subsection{Generalized $CP$ transformations and flavour symmetries}
\label{genCPGf}

We now consider a theory that is invariant under both a flavour symmetry $G_f$ and $CP$. 
We assume $G_f$ to be a discrete and finite group. However, most of the following statements can also be applied to a continuous symmetry $G_f$, if it is global.
An extensive discussion of how to consistently define a $CP$ symmetry in gauge theories can be found in \cite{review_GR}. We assume that
scalar and spinor fields transform according to some representation of the flavour group $G_f$ and we denote a set of fields transforming in a generic irreducible representation ${\bf r}$ of $G_f$  by $\varphi$:
\be
\varphi'=A~ \varphi~~~~,
\label{rho}
\ee 
with $A$ being a unitary matrix depending on the representation ${\bf r}$ and on the chosen group element $g$ of $G_f$. Under $CP$ the multiplet $\varphi$ transforms as
\be
\varphi'= X \varphi^*~~~~.
\label{X2}
\ee
The matrix $X$ is unitary and symmetric and acts on the representation ${\bf r}$.\footnote{It might happen that it is not possible to define the action of $CP$ on a single irreducible representation of $G_f$. In this case $\varphi$ of eq. (\ref{X2}) denotes the smallest combination of irreducible representations of $G_f$ on which the action of $CP$ is well defined.} 

If we perform a $CP$ transformation, followed by a transformation of $G_f$
and another $CP$ transformation we end up with $\varphi'=(X^{-1}A X)^*\varphi$. By consistency this should be a transformation of $G_f$ for the representation ${\bf r}$, given by the matrix $A'$ and representing a group element $g'$ (note in general
$g \neq g'$)
\be
(X^{-1}A X)^*=A'~~~.
\label{Xcond2}
\ee
For a given group $G_f$, this equation constrains the form of $X$ and we would like to determine its general solution that also satisfies eq.~(\ref{Xcond1}). 

As one sees from eqs.(\ref{Xcond1},\ref{Xcond2}), for a solution $X$ also $e^{i \gamma} \, X$ with $\gamma$ being an arbitrary phase is a solution. Notice that at least one solution for $X$ exists and it is the  
 canonical $CP$ transformation $X=\mathbb{1}$, if the representation ${\bf r}$ and also its representation matrices are real, since eqs.(\ref{Xcond1},\ref{Xcond2}) are then trivially fulfilled ($A^*=A=A'$). 
 Since the application of  a similarity transformation and  complex conjugation preserves the usual (matrix) multiplication rules, it is sufficient to check 
 whether a transformation $X$  satisfies the constraint in eq. (\ref{Xcond2}) for a set of generators $A_i$ of the group $G_f$ which give rise, via products, to all elements of $G_f$.
 If $A^n=\mathbb{1}$, $n \in \mathbb{N}$, also $(A')^n=\mathbb{1}$, as can be checked by explicit computation. Thus, $A$ and $A'$ have the same order.
 If $G_f$ is abelian, all irreducible representations are one-dimensional and the constraint in eq. (\ref{Xcond2}) is satisfied by $X=e^{i \gamma}$, with an arbitrary phase $\gamma$. Indeed, $A^*$ is in this case always $A^{-1}$.
If $G_f$ is non-abelian, there are special bases in which we can recognize in a simple way whether there exists (at least) one transformation $X$ which fulfills the constraint in eq. (\ref{Xcond2});
for instance, in a basis in which all non-diagonal generators are real the canonical $CP$ transformation $X=\mathbb{1}$ fulfills eq.(\ref{Xcond2}) for all representation matrices $A$. 
Notice that a generalization of our choice $X$ being symmetric, as mentioned
below eq.(\ref{Xcond1}), can lead to further solutions for $X$. However, in the present paper we always assume $X$ to be symmetric.

If we perform a change of basis with a unitary matrix $\Omega$ in the field space
\be
\tilde{\varphi}=\Omega^\dagger\varphi~~~,
\label{change1}
\ee
the unitary matrices $X$ and $A$ transform as
\be
\tilde{X}=\Omega^\dagger X\Omega^*~~~,~~~~~~~\tilde{A}=\Omega^\dagger A\Omega~~~,
\label{change2}
\ee
as can be seen using eqs. (\ref{rho}, \ref{X2}). The constraints in eqs. (\ref{Xcond1}, \ref{Xcond2}) are covariant under such a transformation $\Omega$, i.e. also $\tilde{X}$ and the set of matrices $\tilde{A}$
fulfill eqs. (\ref{Xcond1},\ref{Xcond2}): $\tilde{X} \tilde{X}^*=\mathbb{1}$ and $(\tilde{X}^{-1} \tilde{A} \tilde{X})^*= \Omega^\dagger A' \Omega=\tilde{A'}$.
A change of basis can be useful in order to reach a basis in which the action of some elements of $G_f$ and/or of $CP$ 
is particularly simple. For example, we can use the result that any unitary symmetric matrix $X$ can be written
as the product $X=\Omega \Omega^T$ of a unitary matrix $\Omega$ and its transpose, in order to go to a basis in which the action of $CP$ is canonical,
$\tilde{X}=\mathbb{1}$, see eq. (\ref{change2}). As a consequence, the constraint in eq. (\ref{Xcond2}) reads $\tilde{A}^*=\tilde{A'}$.

The flavour group $G_f$ together with the $CP$ transformation $X$ defines a group which we call $G_{CP}$ in the following. This group turns out to be a semi-direct product of $G_f$ and $CP$, as we show in appendix \ref{appB}.

\subsection{Lepton mixing from $G_{CP}$}
\label{setup}

One reason for imposing a flavour symmetry $G_f$ is to constrain the form of the lepton mixing matrix in order to explain the observed pattern of mixing angles. A particular approach \cite{Gfnontrivial} is to assume
 the invariance under $G_f$ to be broken in such a way that the combination $m_l^\dagger m_l$ and the neutrino mass matrix $m_\nu$ possess residual discrete symmetries $G_e$ and $G_\nu$, respectively. 
 If we consider three generations of Majorana neutrinos, we are naturally led to the choice $G_\nu=Z_2\times Z_2$,  the largest symmetry of $m_\nu$ leaving neutrino masses unconstrained \cite{S4TB}.  
 For the group $G_e$ we require the following properties: it should be abelian in order to avoid degeneracies among the charged lepton masses and it should allow to assign different charges to the three different generations, for details see \cite{Delta96_384}. 
Thus, $G_e$ is in general a (direct) product of cyclic symmetries, $Z_{m_1} \times \cdot\cdot\cdot \times Z_{m_p}$.
This choice of subgroups of $G_f$ and their relative embedding into the latter predict the form of the lepton mixing matrix, up to permutations of rows and columns and up to arbitrary Majorana phases \cite{S4TB,Delta96_384}.
This mechanism can be implemented in concrete models in which the desired symmetry breaking pattern of the group $G_f$ can be achieved via spontaneous or explicit breaking
and corrections to such a pattern are calculable (and are usually small), see \cite{reviews} for reviews. 

In the present paper we instead consider the case in which the residual symmetry $G_\nu$ is $Z_2\times CP$.
As we will see, this allows us to determine all physical phases and mixing angles in terms of a single real parameter $\theta$. A small non-vanishing mixing angle $\theta_{13}$ can then be
accommodated by a suitable choice of the parameter $\theta$ and furthermore testable relations among the mixing parameters are predicted. 
Once the flavour group $G_f$ has been chosen, several independent definitions of $CP$ are in general possible, leading to physically 
distinct results. In this subsection we explain the setup and show the general form of the lepton mixing matrix, 
while we illustrate several interesting features with an explicit example based on the group $G_f=S_4$ in section \ref{sec3}.

We recall that lepton doublets transform in a three-dimensional irreducible representation ${\bf r}$ of $G_f$ and that neutrinos are of Majorana type. 
We further assume that $G_f$ contains a subgroup $Z_2$ and we denote its generator in the representation ${\bf r}$ by $Z$, $Z^2=\mathbb{1}$.
In order to define the direct product $Z_2\times CP$, $CP$ should commute with $Z_2$. 
This requirement translates into
\be
XZ^*-ZX=0~~~,
\label{xz}
\ee
as can be checked by computing the subsequent action of $Z_2$ and $CP$ and vice versa on the fields $l$ and by requesting the result to be independent of the ordering of these transformations.
Also the condition in eq.(\ref{xz}) is covariant under the change of basis given in eqs. (\ref{change1},\ref{change2}), i.e. $\tilde{X} \tilde{Z}^* -\tilde{Z} \tilde{X}=0$ with $\tilde{Z}=\Omega^\dagger Z \Omega$ holds. Thus,
it is possible to go to a basis in which $Z$ is diagonal and $X$ canonical (again, this can be checked by explicit computation). We indicate this particular basis 
by a hat:
\be
Z=\Omega \hat{Z}\Omega^\dagger~~~,~~~~~~~X=\Omega \Omega^T~~~.
\label{hat}
\ee
Barring the trivial case ($\hat{Z}=\pm \mathbb{1}$), we can assume, without loss of generality, that $\hat{Z}$ is of the form
\be
\hat{Z}=\pm
\begin{pmatrix}
1&0&0\\
0&-1&0\\
0&0&1
\end{pmatrix}~~~.
\ee
In general the conditions of invariance of the neutrino mass matrix $m_\nu$ under $Z_2\times CP$ are\footnote{If we consider the case in which $X$ is not symmetric, the second equality in eq.(\ref{condmnu}) has to be changed into
$X^T m_\nu X= m_\nu^*$. From this equation and the fact that neutrinos are Majorana particles follows that $X X^*$ has to fulfill $(X X^*)^2=\mathbb{1}$. 
Furthermore, the condition in eq.(\ref{xz}) constrains the form of $X$. The admissible non-symmetric $X$ then
have the property that $X X^*$ is proportional to $Z$ (so that the first equality in eq.(\ref{condmnu}) becomes redundant). Most importantly, such an $X$ constrains the neutrino mass matrix $m_\nu$ in such a way that the neutrino
mass spectrum turns out to be partly degenerate which is not compatible with experimental data \cite{global_latest}. This statement holds for any choice of $G_f$ and $Z$ as long as $X$ has to fulfill eq.(\ref{xz}).}
\be
\label{condmnu}
Z^Tm_\nu Z=m_\nu~~~,~~~~~~~X m_\nu X= m_\nu^*~~~.
\ee
By making use of eq. (\ref{hat}) we see that eq.(\ref{condmnu}) takes the form
\be
\hat{Z} (\Omega^T m_\nu \Omega) \hat{Z}=(\Omega^T m_\nu\Omega)~~~,~~~~~~~(\Omega^T m_\nu\Omega)=(\Omega^T m_\nu\Omega)^*~~.
\ee
These conditions are satisfied by
\be
\Omega^T m_\nu\Omega=
\begin{pmatrix}
m_{11}& 0 &m_{13}\\
0&m_{22}&0\\
m_{13}&0&m_{33}
\end{pmatrix}~~~,~~~~~~~m_{ij}=m_{ij}^*~~~.
\ee
The original matrix $m_\nu$ can be diagonalized by
\be
\label{UnuOmRK}
U_\nu^T m_\nu U_\nu= m_{\nu}^{diag}~~~,~~~~~~~U_\nu=\Omega~ R(\theta)~ K~~~,
\ee
where $R(\theta)$ is a rotation matrix  
\be 
R(\theta)=\begin{pmatrix} \cos \theta & 0 & \sin \theta \\ 0 & 1 & 0 \\ - \sin \theta & 0 & \cos \theta \end{pmatrix}~~~,  \qquad \tan 2\theta = \frac{2~ m_{13}}{m_{33} - m_{11}}~~~.
\label{Rtheta}
\ee
The unitary matrix $K$ is diagonal with entries $\pm 1$ and $\pm i$ which encode the CP parity of the neutrino states and it renders $m_{\nu}^{diag}$ positive (semi-)definite.
Note that the fundamental interval of the parameter $\theta$ is $[0,\pi)$, since $R(\theta+\pi)= R(\theta) \, \mbox{diag} (-1,1,-1)$ and the diagonal matrix can be absorbed into the matrix $K$.
This fixes the contribution from the neutrino sector to the lepton mixing up to permutations of the columns, since neutrino masses are unconstrained in the present framework. 

The unitary matrix $U_e$ from the charged lepton sector is determined by requiring invariance of $m_l^\dagger m_l$ under the subgroup $G_e$ which is in general the product of cyclic groups $Z_{m_i}$, $i=1, ..., p$.
Denoting the generator of the cyclic group $Z_{m_i}$ by $Q_i$, the invariance conditions read 
\be
\label{condml}
Q_i^\dagger m_l^\dagger m_l Q_i = m_l^\dagger m_l 
\ee
for all $i$. Since the set of generators $Q_i$ distinguishes between the three generations of charged leptons, the unitary matrix $U_e$ which simultaneously diagonalizes them is determined (up to permutations
of columns and phases of the column vectors)
\be
\label{Qicond}
U_e^\dagger Q_i U_e = \hat{Q}_i
\ee
with $\hat{Q}_i$ diagonal. Plugging eq.(\ref{Qicond}) into eq.(\ref{condml}) shows that $U_e$ also diagonalizes $m^\dagger_l m_l$, i.e. $U_e^\dagger m_l^\dagger m_l U_e= (m_l^\dagger m_l)^{diag}$.

 Finally, we have 
\be
\label{PMNSres}
U_{PMNS}=U_e^\dagger(Q_i)~ \Omega(Z,X)~ R(\theta) ~ K~~~,
\ee
up to permutations of rows and columns. We have spelled out the dependence on the choice of subgroups, which is specified by the 
generators of $G_e$ and $G_{\nu}$, and on the $CP$ transformation, i.e. the set $(Q_i, Z, X)$.
In our approach neutrino and charged lepton masses remain as undetermined parameters and thus the ordering of rows and columns of the PMNS matrix is not fixed. 
For a given set $(Q_i,Z,X)$ all three mixing angles, the Dirac phase and the two Majorana phases are determined in terms of the parameter $\theta$ whose size depends on the neutrino mass matrix elements $m_{ij}$, see eq.(\ref{Rtheta}). 
The Majorana phases are fixed up to the contribution from the matrix $K$ which can only shift the phases by $\pi$. Due to the covariance of all relevant equations, i.e. the independence of the choice of a particular basis, also the
results for the mixing parameters are basis-independent. Furthermore, they also do not depend on the above-mentioned freedom to multiply $X$ with an arbitrary phase $e^{i \gamma}$.

\subsection{Accidental $CP$ symmetries}
\label{CPacc}

We conclude this section with some remarks about possible accidental $CP$ symmetries. 
In particular, we show conditions which have to be necessarily fulfilled by the generators of the subgroups $G_e$ and $G_\nu$, if such an accidental symmetry is present.
Such conditions turn out to be useful for understanding the results  of the example $G_f=S_4$ that we discuss in the next section. 
We have imposed $CP$ conservation in the neutrino, but not in the charged lepton sector. Thus, we might be led to the conclusion that non-trivial Dirac and Majorana phases
are always generated in our approach. Actually this is not the case. Trivial phases are found when the mass matrices $m_l^\dagger m_l$ and $m_\nu$, constrained by our 
choice of $(Q_i, Z, X)$, satisfy the invariance conditions  
\be
\label{invcond2}
Y^* m_l^\dagger m_l Y = (m_l^\dagger m_l )^*~~~, \qquad  \qquad
Y m_\nu Y = m_\nu^*~~~,
\ee
for some unitary symmetric matrix $Y$, to which we refer as an accidental $CP$ symmetry.
Using the results of appendix \ref{appA} together with those of subsection \ref{setup} we can rewrite these conditions in terms of $(Q_i, Z, X)$ instead of the mass matrices $m_l$ and $m_\nu$.
As one can see, $Y$ satisfies the first equality of eq. (\ref{invcond2}) if and only if 
\be
\label{XQcond}
Q_i Y - Y Q_i^T =0
\ee
for all $i$. This condition ensures that $Y$ is diagonal in the same basis as $Q_i$, that is $Y$ is  diagonal in the charged lepton mass basis. 
Notice that if $Y$ satisfies eq.~(\ref{XQcond}) also $Y \prod \limits_{i=1} ^p (Q_i^*)^{n_i}$, $0 \leq n_i \leq m_i$, does so. Similarly, one can check that the second equality of eq. (\ref{invcond2})
implies 
\be
\label{XXZcond}
Z Y - Y Z^* =0 \;\;\; , \;\;\; X Y^* - Y X^* =0~~~.
\ee
The first equality is of the same form as the requirement for having a direct product of $Z_2 \times CP$, see eq.(\ref{xz}), while the second equality states that the two $CP$ transformations
$X$ and $Y$ commute. These conditions are, however, only necessary but not sufficient to ensure that the second equality of eq.(\ref{invcond2}) holds. 

We can distinguish the following cases: $a)$ we cannot find a $CP$ symmetry $Y$ which fulfills eqs.(\ref{invcond2}-\ref{XXZcond}). Then we have to expect non-trivial Dirac and Majorana phases;
$b)$ a $CP$ symmetry $Y$ exists which fulfills these equations and it is furthermore real and diagonal in the neutrino mass basis. Then all CP phases are trivial, see eq.(\ref{phasestriv}) and appendix \ref{appA}; $c)$ we find a
$CP$ transformation $Y$ which fulfills eq.(\ref{XQcond}) and is diagonal in the neutrino mass basis; however, it is not real in this basis, i.e. $Y$ satisfies the first equality in eq.(\ref{XXZcond}), but not the second one.
Then $Y$ is not a $CP$ symmetry of the setup, but it still leaves $m_l^\dagger m_l$ and $m_\nu^\dagger m_\nu$ invariant, as it is the case for Dirac neutrinos, compare eq.(\ref{Diracnu}). Then we know
$J_{CP}=0$ and $\sin\delta=0$. Furthermore, we can show that $U_{PMNS, ij} \, e^{-i (x_i-x^\nu_j)} =  U^*_{PMNS, ij}$, see appendix \ref{appA}, implies that
\be
\label{Majofixed}
|\sin \alpha|= |\sin (x^\nu_1 -x^\nu_2)|~~~,~~~ |\sin \beta|= |\sin (x^\nu_1 -x^\nu_3)|
\ee
with $x^\nu_i$ being the phases of the diagonal entries of $Y$ in the neutrino mass basis.\footnote{Notice that we have to assume that neither the solar nor the reactor mixing angle are $0$ or $\pi/2$.}

In a setup characterized by $(Q_i, Z, X)$ a candidate for an accidental symmetry is $Y = Z X$ which always satisfies eq. (\ref{XXZcond}) and which is real and diagonal in the neutrino mass basis.\footnote{This can also be seen by 
checking that from eqs.(\ref{xz},\ref{condmnu}) follows that $Z X$ satisfies $(Z X) m_\nu (Z X)= m_\nu^*$ and that it fulfills the constraints in eqs.(\ref{Xcond1},\ref{Xcond2}).}

\section{Example $S_4$ and $CP$}
\label{sec3}

We analyze here the mixing patterns originating from the breaking of
$S_4$ and $CP$ to $G_\nu= Z_2 \times CP$ and to $G_e$ being an abelian subgroup of $S_4$. 
 We choose $S_4$, since it is among the smallest discrete groups with an irreducible 
three-dimensional representation and it is well-known to be the smallest symmetry group which leads to TB mixing, if broken in a non-trivial way \cite{S4TB}.
We first present the group $S_4$ in a  basis convenient for our purposes and then discuss in detail the findings of our comprehensive study of the group
$S_4$ and $CP$. 

\subsection{Group theory of $S_4$}
\label{groupS4}

The group $S_4$ can be defined in terms of three generators $S$, $T$ and $U$ \cite{S4generators}  which fulfill the following relations
\begin{eqnarray}
&& S^2 = E \; , \;\; T^3 = E \; , \;\; U^2= E \; ,\\[0mm]
&&  (S T)^3 = E \; , \;\; (S U)^2 = E \; , \;\; (T U)^2 = E \; , \;\;
  (S T U)^4 = E 
\end{eqnarray}
with $E$ being the neutral element of $S_4$.
Note that the generators $S$ and $T$ alone give rise to the group $A_4$. In the following we are only interested in the two irreducible (faithful) three-dimensional
representations, called ${\bf 3}$ and ${\bf 3'}$ in the notation of \cite{S4generators}, to which we assign the three generations of left-handed leptons. We choose real representation
matrices $S$, $T$ and $U$ (called in the same way as the abstract elements of the group $S_4$) for the representation ${\bf 3'}$:
\begin{equation}
S=
\left(
\begin{array}{ccc}
-1&0&0\\
0&1&0\\
0&0&-1
\end{array}
\right)
\;\; , \;\;\;
T=\frac{1}{2}
\left( \begin{array}{ccc}
1& \sqrt{2}& 1\\
\sqrt{2}&0&-\sqrt{2}\\
-1&\sqrt{2}&-1
\end{array}
\right)
\;\; , \;\;\;
U=
\left(
\begin{array}{ccc}
1&0&0\\
0&1&0\\
0&0&-1
\end{array}
\right) \; .
\label{reps4}
\end{equation}
The representation matrices of the other three-dimensional representation ${\bf 3}$ read $S$, $T$ and $-U$.\footnote{\label{footV}
Notice that 
our choice of basis for $S$, $T$ and $U$ is related to the one of \cite{S4generators} with $\bar{S}$, $\bar{T}$ and $\bar{U}$, through the 
unitary transformation $V$
\begin{equation}\nonumber
V= \left( \begin{array}{ccc}
\sqrt{2/3}&1/\sqrt{3}&0\\
-1/\sqrt{6}&1/\sqrt{3}&-i/\sqrt{2}\\
-1/\sqrt{6}&1/\sqrt{3}&i/\sqrt{2}\
\end{array}
\right)
\end{equation}
so that 
\begin{equation}\nonumber
S = V^\dagger \bar{S} V \;\; , \;
T = V^\dagger \bar{T} V \;\; , \;
U = V^\dagger \bar{U} V \; . 
\end{equation}}

The group $S_4$ has $Z_2$, $Z_3$, $Z_4$ and $Z_2 \times Z_2$ as abelian subgroups. The nine $Z_2$ symmetries are generated by the elements
\begin{equation}
\label{Z2gen}
S, T S T^2 S, T^2 S T S, U, U S, U T, U T^2, U S T S, U S T^2 S \, ,
\end{equation}
while the generators of the four $Z_3$ symmetries can be chosen as
\begin{equation}
 T   \; , \;\;
 S T  \; , \;\;
 S T^2 \; , \;\;
 T S T  \; ,
\end{equation}
and those of the three $Z_4$ symmetries as
\begin{equation}
 S T U  \; , \;\;
 U T S  \; , \;\;
 U T^2 S T  \; .
\end{equation}
The $Z_2$ generating elements are divided into two classes (the first three of the list in eq.(\ref{Z2gen}) and the last six ones), while the $Z_3$ subgroups as well as the $Z_4$ subgroups of $S_4$ are all conjugate to each other.
There are four Klein subgroups. One of them, called $K_N$, is normal, while the three other ones $K_i$, $i=1,2,3$ are conjugate to each other. Possible
sets of generators of the different Klein subgroups are
\begin{equation}
K_N \, : \; S, T S T^2 S \;\; , \;\;\;
K_1 \, : \; S, U \;\; , \;\;\;
K_2 \, : \; T S T^2 S, U T^2 \;\; , \;\;\;
K_3 \, : \; T^2 S T S, U T \;\; . \;\;\;
\end{equation}

\subsection{Mixing patterns from flavour groups $S_4$ and $A_4$ without $CP$}
\label{A4S4Gfpatterns}

We briefly repeat the mixing patterns which can be derived from the group $S_4$ as well as $A_4$, if the group $G_\nu$ is a Klein group
and $G_e$ an abelian subgroup of $G_f$, capable to distinguish the three generations of charged leptons. All these results can be found
in \cite{A4S4patterns, Delta96_384}. We have three different viable choices for the group $G_e$ if $G_f=S_4$: $G_e=Z_3$, $G_e=Z_4$
and $G_e=Z_2 \times Z_2$. As is well-known, the choice $G_e=Z_3$ leads to TB mixing \cite{TB} (for example, we can choose $T$ as generator for
$G_e$ and $S$ and $U$ for $G_\nu$), while $G_e=Z_4$ (one possible choice of generator is $S T U$) and $G_e=Z_2 \times Z_2$
(for example generated by the elements $T S T^2 S$ and $U T^2$) give rise to bimaximal (BM) mixing \cite{BM}. In the case of $G_f=A_4$, we have a unique 
choice for $G_\nu=Z_2 \times Z_2$ (to be generated by $S$ and $T S T^2 S$) and as choice for $G_e$ only $G_e=Z_3$ (for example, generated
by $T$). The mixing pattern is given by the familiar democratic mixing matrix, in which all elements have the same absolute value and the mixing parameters
read $\sin^2 \theta_{13}=1/3$, $\sin^2 \theta_{12}=1/2$, $\sin^2 \theta_{23}=1/2$, $|J_{CP}|=1/(6 \sqrt{3})$ and $|\sin \delta|=1$.
This matrix has already been discussed many years ago as possible lepton mixing matrix \cite{A4magic}.

\subsection{Results for $G_f=S_4$ and $CP$}
\label{resS4}

We show the results of a comprehensive study in which we assume $G_f=S_4$, $G_\nu=Z_2 \times CP$ and $G_e$  being $G_e= Z_3$, $G_e=Z_4$
or $G_e= Z_2 \times Z_2$. In order to facilitate understanding we present our results in terms of examples for the different cases.  
We find that it is sufficient to consider only a small number of cases which lead to different results for mixing angles and CP phases, since other possible
 choices of $Q_i$, $Z$ and $X$ are related by similarity transformations -belonging to the group $S_4$- to our representative solutions and thus cannot lead
 to new results.
We concentrate in our discussion on the representation ${\bf 3^\prime}$ of $S_4$. However, if we assigned the three generations of left-handed leptons to the representation 
${\bf 3}$ instead, the results would be the same and no additional results would be found, because the generators of the triplet ${\bf 3}$
just differ in the overall sign of the generator $U$ from those of the representation ${\bf 3^\prime}$.
It follows a short subsection with technical details necessary for the derivation of the results.

\subsubsection{Choice of $(Q_i, Z, X)$}
\label{technical}

All possible choices of $Z$ and $X$ are related through similarity transformations (contained in $S_4$) to the following three $Z$ 
\begin{equation}
Z=S \; , \;\; Z=S U \;\;\; \mbox{and} \;\;\; Z=U 
\end{equation}
and their corresponding $X_i$ which fulfill the requirements stated in eqs.(\ref{Xcond1},\ref{Xcond2},\ref{xz}) of section \ref{sec2}. For all $Z$
\begin{eqnarray}\nonumber
&&X_1=\mathbb{1}
\; , \;\;
X_2=\left(
\begin{array}{ccc}
-1&0&0\\
0&1&0\\
0&0&-1
\end{array}
\right)  \; , \;\;
X_3=\left(
\begin{array}{ccc}
1&0&0\\
0&1&0\\
0&0&-1
\end{array}
\right) \; , \;\;
\\
&& X_4=\left(
\begin{array}{ccc}
-1&0&0\\
0&1&0\\
0&0&1
\end{array}
\right)
\label{admissX_1}
\end{eqnarray}
are admissible and for $Z=S$ in addition 
\begin{equation}
X_5=\left(
\begin{array}{ccc}
0&0&-1\\
0&-1&0\\
-1&0&0
\end{array}
\right)  \; , \;\;
X_6=\left(
\begin{array}{ccc}
0&0&1\\
0&-1&0\\
1&0&0
\end{array}
\right)  \; .
\label{admissX_2}
\end{equation}
As mentioned in section \ref{sec2}, these transformations are defined up to an overall phase. $X_1$ is the canonical $CP$ transformation.
Notice that in our particular basis the different $X_i$ turn out to be proportional to elements of the group $S_4$:
$X_2 \propto S$, $X_3 \propto U$, $X_4 \propto S U$, $X_5 \propto T S T^2 S$ and $X_6 \propto T^2 S T S$.
However, this is just a coincidence and in general the transformations $X_i$ do not belong to the
flavour group $G_f$. 

We list the transformations $\Omega_i$  which bring the different $X_i$ into the canonical form $\tilde X =\mathbb{1}$, see eqs.(\ref{change2},\ref{hat}):
\begin{eqnarray}\nonumber
&& \Omega_{1}=\mathbb{1}
\; , \;\;
\Omega_{2}=\left(
\begin{array}{ccc}
i&0&0\\
0&1&0\\
0&0&i
\end{array}
\right) \; , \;\;
\Omega_{3}=\left(
\begin{array}{ccc}
1&0&0\\
0&1&0\\
0&0&i
\end{array}
\right) \; , \;\;
\Omega_{4}=\left(
\begin{array}{ccc}
i&0&0\\
0&1&0\\
0&0&1
\end{array}
\right) \; , \;\;
\\ 
&& 
\Omega_{5}=\frac{1}{\sqrt{2}}
\left( \begin{array}{ccc}
-1&0&i\\
0&\sqrt{2} \, i&0\\
1&0&i
\end{array}
\right) \; , \;\;
\Omega_{6}=\frac{1}{\sqrt{2}}
\left( \begin{array}{ccc}
-i&0&1\\
0&\sqrt{2} \, i&0\\
i&0&1
\end{array}
\right) \; .
\end{eqnarray}
We note that, obviously, for different choices of $Z$, but using the same form of the matrices $\Omega_i$, the rotation matrix $R (\theta)$ defined in eq. (\ref{Rtheta}) changes its form:
 it is a rotation in the (13)-plane for $Z=S$, a rotation in the (23)-plane for $Z=S U$ and for $Z=U$ a rotation in the (12)-plane.

The different choices of $G_e$, for which we discuss lepton mixing, can be represented by 
\be
\begin{array}{lll} 
Q=T &\mbox{for}& G_e=Z_3 \; ,
\\  
Q=S T U &\mbox{for}&  G_e=Z_4 \; ,
\\ 
Q_1=T S T^2 S~,~~ Q_2=U T^2 &\mbox{for}& G_e=Z_2 \times Z_2 \; . 
\end{array}
\ee
The matrix $U_e$ which diagonalizes the charged lepton mass matrix $m_l^\dagger m_l$ is then of
the form
\begin{equation}
\label{UeQT}
U_e = \left( \begin{array}{ccc}
  \sqrt{2/3} & -1/\sqrt{6} & -1/\sqrt{6}\\
  1/\sqrt{3} & 1/\sqrt{3} & 1/\sqrt{3}\\
  0 & i/\sqrt{2} & -i/\sqrt{2}
\end{array}
\right)
\;\;\; \mbox{for} \; Q=T \; ,
\end{equation}
\begin{equation}
\label{UeQSTU}
U_e = \left( \begin{array}{ccc}
-1/\sqrt{2} & 1/2 & 1/2\\
0 & -i/\sqrt{2} & i/\sqrt{2}\\
1/\sqrt{2} & 1/2 & 1/2
\end{array}
\right)
\;\;\; \mbox{for} \; Q=S T U
\end{equation}
and
\begin{equation}
\label{UeQ1Q2}
U_e = \left( \begin{array}{ccc}
 1/2 & -1/\sqrt{2} & 1/2\\
- 1/\sqrt{2} & 0 & 1/\sqrt{2}\\
 1/2 & 1/\sqrt{2} & 1/2
\end{array}
\right)
\;\;\; \mbox{for} \; Q_1=T S T^2 S \; \mbox{and} \; Q_2=U T^2 \; .
\end{equation}

\subsubsection{Lepton mixing parameters}
\label{mixing}

We have performed a comprehensive study in which we consider
all possible choices of $Q_i$, $Z$ and $X$ and all possible permutations of columns and rows of the mixing matrix. However, we show
in the following only results which we consider phenomenologically interesting
 in the sense that we can find a value of the parameter $\theta$ such that the resulting mixing angles are reasonably close to their best fit values which we take from \cite{global_latest}. As measure we
 use a $\chi^2$ function defined in the usual way and require its minimal value to be less than $100$.  In this way, all solutions leading to vanishing $\theta_{13}$ independent of the parameter $\theta$ are excluded, since $\theta_{13}=0$ 
 is disfavoured at the $10 \, \sigma$ level by global fits, $\sin^2 \theta_{13}= 0.023 \pm 0.0023$, \cite{global_latest}.
 We define two different $\chi^2$ functions, because $\sin^2 \theta_{23}$ has two best fit values
 $\sin^2 \theta_{23}=0.41$ and $\sin^2 \theta_{23}=0.59$.\footnote{Note that the $1 \sigma$ errors are not completely gaussian for $\sin^2 \theta_{23}$. However, we use for the smaller best fit value a $1 \sigma$ error of $\pm 0.031$ and for the larger one an error of $\pm 0.022$.} For this reason, we display in the 
 tables below in several occasions different best fit values $\theta_{\mbox{bf}}$ for the parameter $\theta$ for which the $\chi^2$ function has a global minimum.  
 In the tables we also display the results for the sines of the CP phases $\delta$, $\alpha$ and $\beta$ (and the Jarlskog invariant $J_{CP}$). 
 These quantities are presented in terms of absolute values, since the sign of the Jarlskog invariant $J_{CP}$ depends on the ordering of rows and columns, while the sign of $\sin\alpha$ and
 $\sin\beta$ depends on the CP parity of the neutrino states which is encoded in the matrix $K$, see eq.(\ref{UnuOmRK}) in section \ref{sec2} (changing CP parity shifts the Majorana phase by $\pi$).
 
 Requiring $\chi^2 < 100$, we find five viable solutions (Case I, II, IV, V and the case with $G_e=Z_4$ or $G_e=Z_2 \times Z_2$). 
 However, in all these cases the Majorana phases are trivial, i.e. $\sin \alpha=0$ and $\sin \beta=0$. Thus, we have included another case,  called Case III, which leads to non-trivial
 Majorana phases depending on the parameter $\theta$, although the minimum value of its $\chi^2$ functions is above 100.
 
In the following we put special emphasis on the results for the CP phases $\delta$, $\alpha$ and $\beta$, because frequently an accidental $CP$ symmetry
 is present which leads to a trivial Dirac phase and/or Majorana phases.
 
 We first take $G_e=Z_3$ and choose $Q=T$. Five different cases can be distinguished and are represented by
 \be
 \begin{array}{ccc}
 \mbox{I} & Z=S \; ,& X=X_1\\
  \mbox{II} & Z=S \; ,& X=X_3\\
  \mbox{III} & Z=S \; ,& X= X_5\\
  \mbox{IV} & Z=S U \; ,& X= X_1\\
  \mbox{V} & Z=S U \; ,& X=X_2 \; .
\end{array}
 \ee
 Notice that in Case I $X_2$ is also a $CP$ symmetry of the neutrino sector, since $Z X_1=X_2$. Similarly, it holds in Case II that $X_4=Z X_3$ is also a $CP$ symmetry and
 in Case III it is $X_6=Z X_5$. Analogously, we find that in Case IV $X_4=Z X_1$ is also present and in Case V $X_3=Z X_2$. Thus, the six possible choices of $X_i$ for $Z=S$, mentioned in subsection \ref{technical},
 give rise to three independent cases and for $Z=S U$ the four possible $X_i$ lead to effectively two different cases.
In Cases I, II, IV and V we find a value of $\theta$ for which the computed mixing angles agree rather well with the ones obtained in global fits \cite{global_latest}. This is shown in table \ref{tab_c1_c5}
together with the formulae for the mixing parameters in terms of generic $\theta$.

\begin{table}[t!]
\begin{center}
 \begin{tabular}{|c|c|c|c|c|} 
\hline
 &I & II & IV & V\\
\hline
$\sin^2 \theta_{13}$ & $\frac{2}{3} \sin^2 \theta$ & $\frac{2}{3} \sin^2 \theta$ & $\frac{1}{3} \sin^2 \theta$& $\frac{1}{3} \sin^2 \theta$ \\
 $\sin^2 \theta_{12}$ & $\frac{1}{2+\cos 2 \theta}$ & $ \frac{1}{2+\cos 2 \theta}$ & $\frac{\cos^2 \theta}{2+\cos^2 \theta}$ & $\frac{\cos^2 \theta}{2+\cos^2 \theta}$\\
 $ \sin^2 \theta_{23}$ & $\frac{1}{2}$ & $\frac{1}{2} \, \left( 1-\frac{\sqrt{3} \sin 2 \theta}{2 +\cos 2\theta}\right)$ & $\frac{1}{2}$ & $\frac{1}{2} \, \left( 1- \frac{2 \sqrt{6} \sin 2 \theta}{5+\cos 2\theta} \right)$\\
$|J_{CP}|$ & $\frac{|\sin 2 \theta|}{6 \sqrt{3}}$ & 0 & $\frac{|\sin 2\theta|}{6 \sqrt{6}}$ & 0\\
$|\sin \delta|$ & $1^{a}$ & 0 & $1^{a}$ &  0\\
$\sin \alpha$ & 0 & 0 & 0 & 0\\
$\sin \beta$ & 0 & 0 & 0 & 0\\
\hline
 $\theta_{\mbox{bf}}$ & 0.185& $\begin{array}{cc} 0.184, & \theta_{23} < \pi/4\\ 2.958, & \theta_{23} > \pi/4 \end{array}$ & 0.268 & $\begin{array}{cc} 0.251, & \theta_{23} < \pi/4\\ 2.907, & \theta_{23} > \pi/4 \end{array}$\\
$\chi^2_{\mbox{min}}$ &  $\begin{array}{cc} 18.4, & \theta_{23} < \pi/4\\ 26.7, & \theta_{23} > \pi/4 \end{array}$ &
 $\begin{array}{cc} 10.3, & \theta_{23} < \pi/4\\ 10.5, & \theta_{23} > \pi/4 \end{array}$ & $\begin{array}{cc} 10.2, & \theta_{23} < \pi/4\\ 18.5, & \theta_{23} > \pi/4 \end{array}$ 
 & $\begin{array}{cc} 16.1, & \theta_{23} < \pi/4\\ 27.2, & \theta_{23} > \pi/4 \end{array}$\\
\hline
$\sin^2 \theta_{13} (\theta_{\mbox{bf}})$ & 0.023 & 0.022 & 0.023 & $\begin{array}{cc} 0.021, & \theta_{23} < \pi/4\\ 0.018, & \theta_{23} > \pi/4 \end{array}$\\ 
$\sin^2 \theta_{12} (\theta_{\mbox{bf}})$ & 0.341 & 0.341 & 0.317 & $\begin{array}{cc} 0.319, & \theta_{23} < \pi/4\\  0.321, & \theta_{23} > \pi/4 \end{array}$\\ 
$\sin^2 \theta_{23} (\theta_{\mbox{bf}})$ & 0.5 & $\begin{array}{cc} 0.394, & \theta_{23} < \pi/4\\  0.606, & \theta_{23} > \pi/4 \end{array}$ & 0.5 
& $\begin{array}{cc} 0.299, & \theta_{23} < \pi/4\\ 0.688, & \theta_{23} > \pi/4 \end{array}$ \\ 
$|J_{CP}| (\theta_{\mbox{bf}})$ & 0.0348 & 0 & 0.0348 & 0\\ 
\hline
 \end{tabular}
 \end{center}
\begin{center}
\caption{\small Results for the mixing parameters in terms of the parameter $\theta$ for the Cases I, II, IV and V. We display the best fit value $\theta_{\mbox{bf}}$ for $\theta$
for which the $\chi^2$ functions have a global minimum $\chi^2_{\mbox{min}}$. Since the global fit \cite{global_latest} gives two best fit values for $\sin^2 \theta_{23}$, one 
with $\theta_{23} < \pi/4$ and one with $\theta_{23} > \pi/4$, we distinguish these two possibilities. Since the CP phases are independent of $\theta$, we only
give the values of the mixing angles and the absolute value of $J_{CP}$ for $\theta=\theta_{\mbox{bf}}$. \newline
$^a$ In the special case $\sin 2 \theta =0$ $\sin \delta$ vanishes, since then $J_{CP}$ vanishes. However, then the mixing angles are in considerable disagreement with the best fit values
from \cite{global_latest}. \label{tab_c1_c5} \normalsize}
 \end{center}
\end{table}
In Case II and Case V all CP phases are trivial. Thus,  an accidental $CP$ symmetry $Y$ common to the charged lepton and neutrino sector has to be present.
Indeed, $Y=X_3$ is such an accidental symmetry in both cases, because it satisfies eq.(\ref{XQcond}) for $Q=T$ and, in the neutrino sector, it is once imposed as $X=X_3$ (Case II) and once not directly imposed, 
but $Z X_2=X_3$ holds (Case V) [implying in both cases that the conditions in eq.(\ref{XXZcond}) are fulfilled and $Y$ is diagonal and real in the neutrino mass basis].
\begin{figure}[t!]
 \parbox{6.5in}{\includegraphics[scale=0.8]{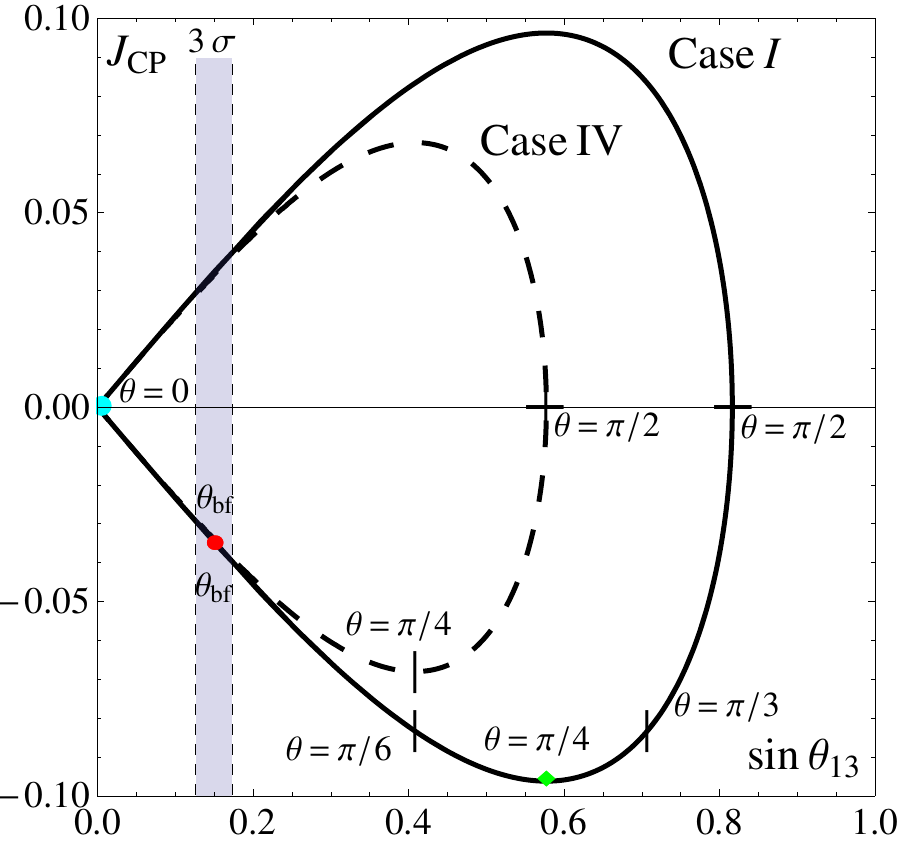}
 \includegraphics[scale=0.8]{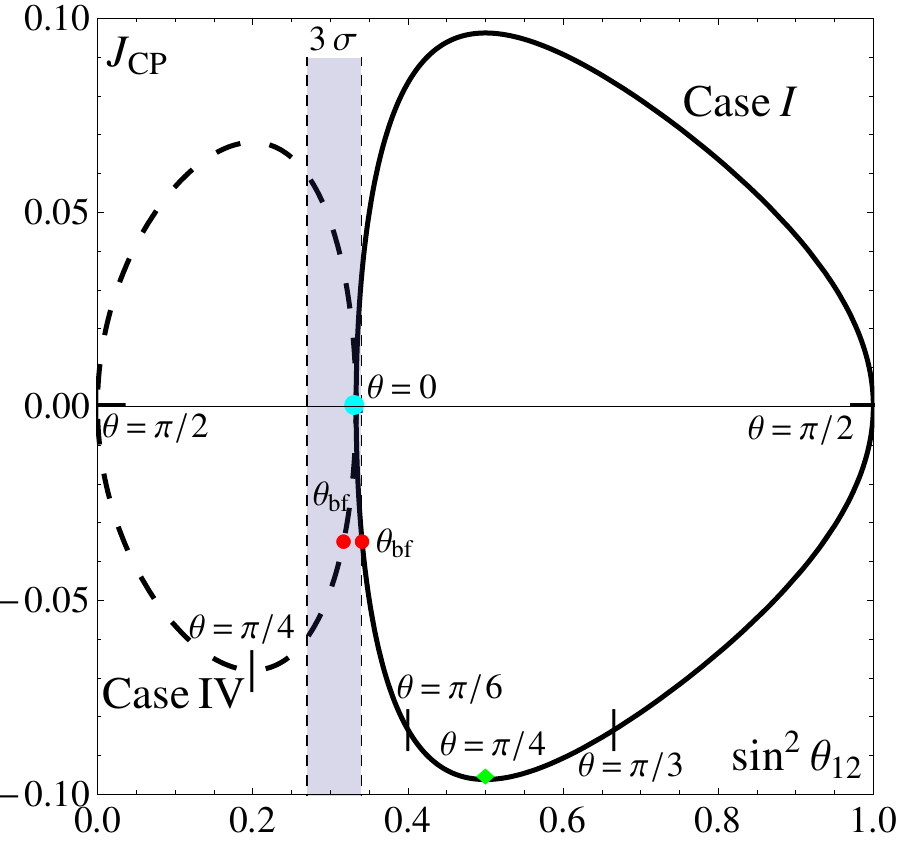}\\
$\phantom{X}$\includegraphics[scale=0.8]{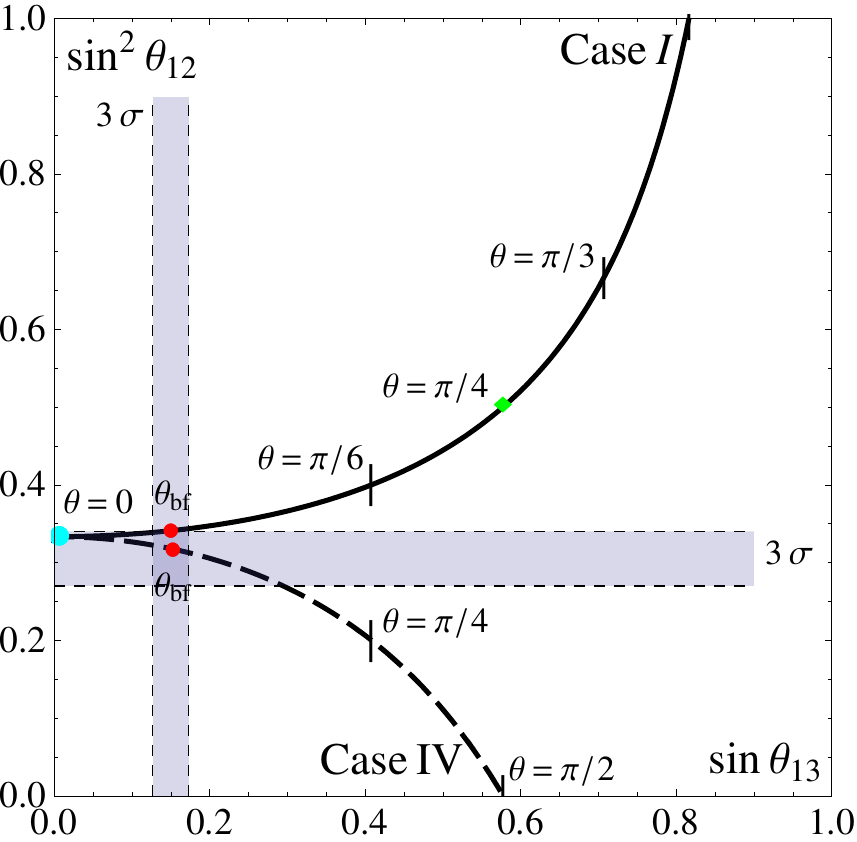} \parbox{3in}{\vspace{-2in} {Figure 1: \small Results for the mixing parameters $\sin \theta_{13}$, $\sin^2 \theta_{12}$
 and $J_{CP}$ for Case I (straight line) and Case IV (dashed line). We mark the 
 value $\theta_{\mbox{bf}}$ of the parameter $\theta$ for which the $\chi^2$ functions have a global minimum with a red dot. We also indicate special points
 $\theta=0$ and $\theta=\pi/4$ for which the second $Z_2$ symmetry in the neutrino sector becomes part of the group $S_4$ and well-known mixing patterns like TB mixing and the democratic
 mixing are reproduced. For better guidance of the eye we also mark $\theta=\pi/n$, $n=6,4,3,2$ on the curves. The shown $3 \, \sigma$ ranges for the mixing angles are taken from \cite{global_latest}.
\normalsize}}}
\end{figure}

In Case I and Case IV $\sin\delta$ vanishes for $\sin 2\theta=0$ and the Majorana phases are always trivial. 
This result can be traced back to the facts that $X_3$ is a $CP$ symmetry in the charged lepton sector, that it fulfills the conditions in eq.(\ref{XXZcond}) and that its form in the neutrino mass basis is
\begin{equation}
Y_\nu = U^\dagger_\nu X_3 U_\nu^\star \;\;\; \mbox{with} \;\;\; U_\nu = \Omega_1 R(\theta) K \; ,
\end{equation}
which explicitly reads for Case I and Case IV
\begin{equation}
Y_{\nu, \mbox{\tiny I}\normalsize} = K^\star \left(
\begin{array}{ccc}
 \cos 2 \theta & 0 & \sin 2 \theta \\
 0 & 1 & 0\\
\sin 2 \theta & 0 & -\cos 2 \theta
\end{array}
\right) K^\star
\;\;\; \mbox{and} \;\;\;
Y_{\nu, \mbox{\tiny IV} \normalsize} = K^\star \left(
\begin{array}{ccc}
 1 & 0 & 0\\
 0 & \cos 2 \theta & \sin 2 \theta\\
 0 & \sin 2 \theta & -\cos 2 \theta
\end{array}
\right) K^\star
\; ,
\end{equation}
respectively. For the special choice $\sin 2 \theta =0$ (and thus $\cos 2 \theta = \pm 1$) these
matrices are diagonal and real and thus $X_3$ is an accidental $CP$ symmetry of the charged lepton and neutrino mass matrices
and consequently all CP phases are trivial.
\begin{figure}[t!]
\parbox{6.5in}{\includegraphics[scale=0.8]{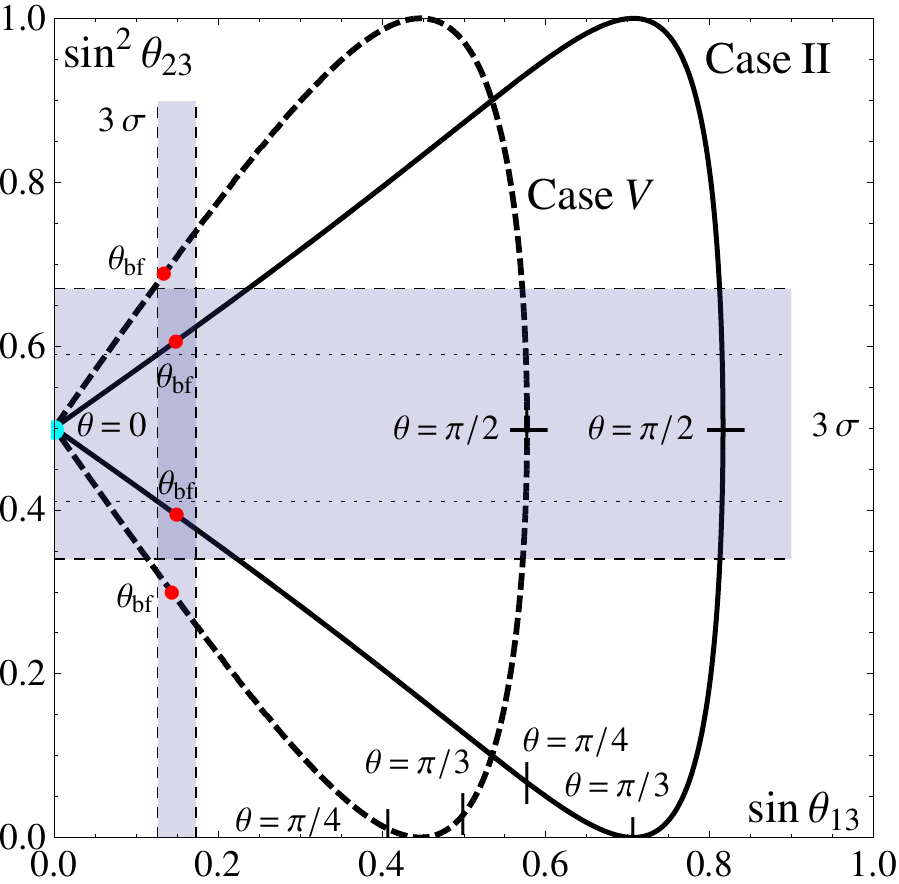}
 \includegraphics[scale=0.8]{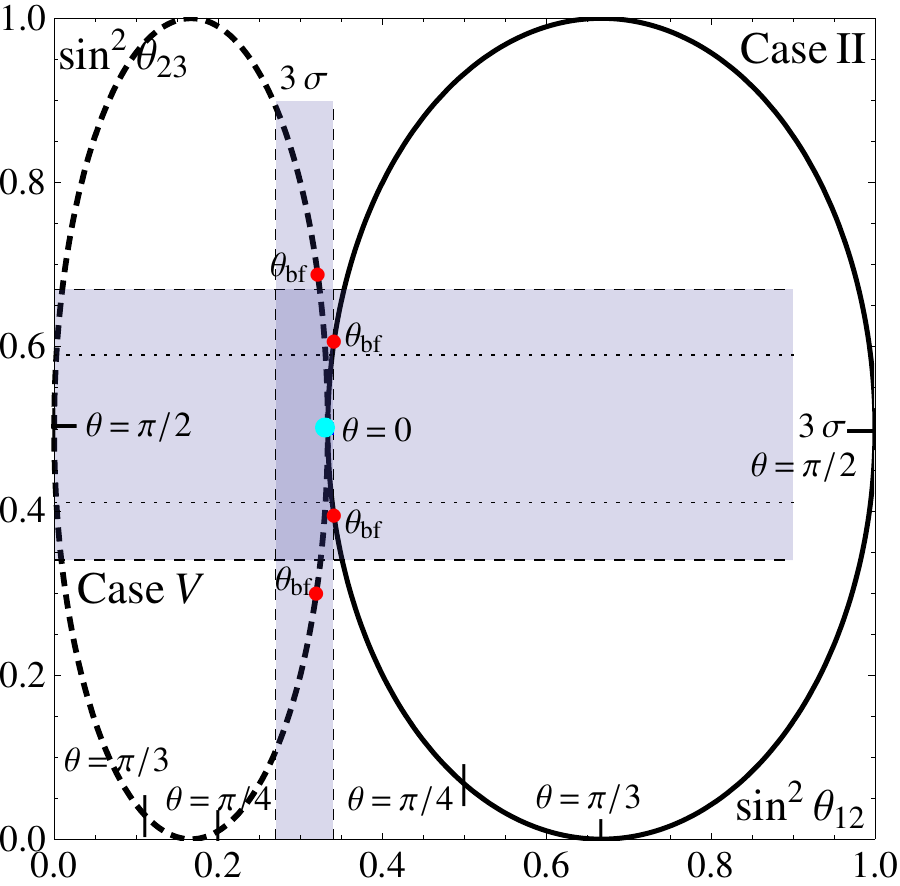}\\
Figure 2: \small Results for the atmospheric, solar and reactor mixing angles for Case II (straight line) and Case V (dashed line). We mark the 
 value $\theta_{\mbox{bf}}$ of the parameter $\theta$ for which the $\chi^2$ functions have a global minimum with a red dot. We also indicate the special point
 $\theta=0$ for which the second $Z_2$ symmetry in the neutrino sector becomes part of the group $S_4$ and TB mixing is reproduced. 
 For better guidance of the eye we also mark $\theta=\pi/n$, $n=4,3,2$ on the curves. The shown $3 \, \sigma$ ranges for the mixing angles and the best fit values of the atmospheric mixing angle 
 are taken from \cite{global_latest}. Notice that the plot for 
 $\sin^2 \theta_{12}$ and $\sin \theta_{13}$ is the same as for Case I and Case IV and can be found in figure 1.
 \normalsize}
 \end{figure}

As regards the mixing angles, the atmospheric mixing turns out to be maximal in Case I and Case IV, because the elements  
of the second and third rows of the PMNS matrix have the same absolute values, i.e. $|U_{\mu i}|=|U_{\tau i}|$ for $i=1,2,3$ \cite{mutaureflection_HS,mutaureflection_GL,max_CPV_D27}.
This feature also explains the maximal Dirac phase (as long as $\sin\theta_{13} \neq 0$ and $\sin 2 \theta_{12} \neq 0$).
A change of basis (using $\Omega=V^\dagger$, see footnote \ref{footV}) shows that the prediction of the atmospheric mixing angle as well as the Dirac phase are attributed to
imposing as one of the symmetries of the neutrino mass matrix the so-called $\mu\tau$ reflection symmetry \cite{mutaureflection_HS,mutaureflection_GL}, 
while the charged lepton mass matrix is diagonal.\footnote{The other symmetry in the neutrino sector is a $Z_2$ symmetry generated by 
\be\nonumber
\tilde{Z}= \frac 13 \, \left( \begin{array}{ccc}
 -1 & 2 & 2\\
 2 & -1 & 2\\
 2 & 2 & -1
\end{array}
\right) \;\;\; \mbox{and} \;\;\;
\tilde{Z}= \frac 13 \, \left( \begin{array}{ccc}
 -1 & 2 & 2\\
  2 & 2 & -1\\
 2 & -1 & 2
\end{array}
\right)
\ee
in Case I and Case IV, respectively.}
In Case I and Case II  $Z=S$ enforces the second column of the PMNS matrix to be tri-maximal. As a consequence, the solar mixing angle has a lower limit given by $\sin^2 \theta_{12} \geq 1/3$ \cite{TM_pheno}, which is  disfavoured by the global fits \cite{global_latest}, $\sin^2 \theta_{12}= 0.30 \pm 0.013$, at the $2 \, \sigma$ level.
On the other hand, in Case IV and Case V $Z=S U$ leads to a lepton mixing matrix whose first column coincides with the one of TB mixing, up to a possible phase, and consequently the solar mixing angle has an upper limit
given by $\sin^2 \theta_{12} \leq 1/3$ \cite{TB_first_column}. The results of Case I and Case II have been discussed previously in the literature 
as generalization of TB mixing  \cite{mutaureflection_HS} (the mixing patterns were named tri$\chi$maximal for Case I and tri$\phi$maximal for Case II, respectively). 
The mixing parameters in the different cases are illustrated in figures 1 and 2. 
For a better presentation we display $\sin \theta_{13}$ instead of its square.\footnote{We take as best fit value of $\sin \theta_{13}$ $\sin \theta_{13}=0.15$ and as $3 \, \sigma$ range 
$0.126 \leq \sin \theta_{13} \leq 0.173$. These values are derived from \cite{global_latest}.}
 We show 
the reactor and the solar mixing angles and $J_{CP}$ for Case I and Case IV, while we present the three mixing angles for Case II and Case V. Notice that the plot in the plane 
of $\sin^2 \theta_{12}$ and $\sin \theta_{13}$ is the same in Case I (IV) and Case II (V) . Furthermore, we mark the best fit value $\theta_{\mbox{bf}}$ for which the $\chi^2$ functions 
have a global minimum with a red dot. For better guidance of the eye we also indicate the values $\theta=\pi/n$, $n=6,4,3,2$ on the curves.
  
\begin{table}[t!]
\begin{center}
 \begin{tabular}{|c|c|c|} 
\hline
&\multicolumn{2}{|c|}{III}\\
\hline
$\sin^2 \theta_{13}$ & \multicolumn{2}{|c|}{$\frac{1}{3} \left( 1 - \frac{\sqrt{3}}{2} \sin 2 \theta \right)$}\\
 $\sin^2 \theta_{12}$ & \multicolumn{2}{|c|}{$\frac{2}{4 + \sqrt{3} \sin 2 \theta}$}\\
 $ \sin^2 \theta_{23}$ &$\frac{2}{4 +\sqrt{3} \sin 2 \theta}$ & $1-\frac{2}{4 +\sqrt{3} \sin 2 \theta}$\\
$|J_{CP}|$ &  \multicolumn{2}{|c|}{$\frac{|\cos 2 \theta|}{6 \sqrt{3}}$}\\
$|\sin \delta|$ &  \multicolumn{2}{|c|}{$\left|\frac{(4 + \sqrt{3} \sin 2 \theta) \cos 2 \theta \sqrt{4-2\sqrt{3} \sin 2 \theta}}{5+3 \cos 4 \theta}\right|$}\\
$|\sin \alpha|$ &  \multicolumn{2}{|c|}{$\left|\frac{\sqrt{3}+2 \sin 2 \theta}{2+\sqrt{3} \sin 2 \theta}\right|$}\\
$|\sin \beta|$ &  \multicolumn{2}{|c|}{$\left|\frac{4 \sqrt{3} \cos 2 \theta}{5+3 \cos 4 \theta}\right|$}\\
\hline
 $\theta_{\mbox{bf}}$ & 0.785, $\, \theta_{23} < \pi/4$ & 0.785, $\, \theta_{23} > \pi/4$\\
$\chi^2_{\mbox{min}}$ & 106.7 & 110.5\\
\hline
$\sin^2 \theta_{13} (\theta_{\mbox{bf}})$ &   \multicolumn{2}{|c|}{0.045}\\ 
$\sin^2 \theta_{12} (\theta_{\mbox{bf}})$ &   \multicolumn{2}{|c|}{0.349} \\ 
$\sin^2 \theta_{23} (\theta_{\mbox{bf}})$ & 0.349 & 0.651 \\ 
$|J_{CP}| (\theta_{\mbox{bf}})$ &   \multicolumn{2}{|c|}{0} \\ 
$|\sin \delta| (\theta_{\mbox{bf}})$  & \multicolumn{2}{|c|}{0}\\
$|\sin \alpha| (\theta_{\mbox{bf}})$ &   \multicolumn{2}{|c|}{1}\\
$|\sin \beta| (\theta_{\mbox{bf}})$ &   \multicolumn{2}{|c|}{0} \\
\hline
 \end{tabular}
 \end{center}
 \begin{center}
\caption{\small Results for the mixing parameters in Case III. As one can see this is the only case with a non-trivial dependence of the CP phases on the parameter $\theta$. The possibility to exchange the second and
third rows of the PMNS matrix gives rise to the two solutions which differ in their result for $\sin^2 \theta_{23}$. 
The $\chi^2$ functions have a global minimum with $\chi^2 \gtrsim 100$ at $\theta_{\mbox{bf}} \approx \pi/4$. 
\label{tab_c3} \normalsize}
\end{center}
\end{table}
As we have mentioned, we discuss Case III because it is the only one in which all CP phases are in general non-trivial and depend on the parameter $\theta$, although in this
case the minimum value of the $\chi^2$ functions is (slightly) larger than 100. The results are collected in table \ref{tab_c3}.
Again, $Z=S$ is responsible for the fact that the second column of the PMNS matrix is tri-maximal and thus the solar mixing angle has a lower limit $\sin^2\theta_{12} \geq 1/3$.
Furthermore, we note that the solar and the atmospheric mixing angles are closely related because their sine squares are either equal or fulfill  $\sin^2 \theta_{12}= 1-
\sin^2 \theta_{23}$.

 \begin{table}[t!]
 \begin{center}
  \begin{tabular}{|c|c|c|} 
\hline
&\multicolumn{2}{|c|}{$G_e=Z_4$ or $G_e=Z_2 \times Z_2$}\\
\cline{2-3}
& Case a& Case b\\
\hline
$\sin^2 \theta_{13}$ & \multicolumn{2}{|c|}{$\frac{1}{4} \left( \sqrt{2} \cos \theta + \sin \theta \right)^2$}\\
 $\sin^2 \theta_{12}$ & \multicolumn{2}{|c|}{$\frac{2}{5 - \cos 2 \theta -2  \sqrt{2} \sin 2 \theta}$}\\
 $ \sin^2 \theta_{23}$ &$\frac{4 \sin^2 \theta}{5 - \cos 2 \theta - 2 \sqrt{2} \sin 2 \theta}$ & $1-\frac{4 \sin^2 \theta}{5 - \cos 2 \theta - 2 \sqrt{2} \sin 2 \theta}$\\
$J_{CP}$ &  \multicolumn{2}{|c|}{$0$}\\
$\sin \alpha$ & \multicolumn{2}{|c|}{$0$}\\
$\sin \beta$ & \multicolumn{2}{|c|}{$0$}\\
\hline
 $\theta_{\mbox{bf}}$ & 2.009, $\, \theta_{23} < \pi/4$ & 2.010, $\, \theta_{23} > \pi/4$\\
$\chi^2_{\mbox{min}}$ & 11.6 & 11.7\\
\hline
$\sin^2 \theta_{13} (\theta_{\mbox{bf}})$ & \multicolumn{2}{|c|}{0.023} \\ 
$\sin^2 \theta_{12} (\theta_{\mbox{bf}})$ & \multicolumn{2}{|c|}{0.256} \\ 
$\sin^2 \theta_{23} (\theta_{\mbox{bf}})$ & 0.420 & 0.581 \\ 
\hline
 \end{tabular}
 \end{center}
  \begin{center}
 \caption{\small Results for the mixing parameters in terms of $\theta$ for the only viable case which we can find for $G_e=Z_4$ or $G_e=Z_2 \times Z_2$. The possible exchange of the second
 and third rows of the PMNS matrix gives rise to the 
 two different solutions, Case a and Case b. All CP phases are trivial independently of the parameter $\theta$. We also display the mixing parameters at $\theta_{\mbox{bf}}$, best fit points
 for which the $\chi^2$ functions have a global minimum.
 \label{tab_Z4}
 \normalsize}
  \end{center}
 \end{table}
The best fit value $\theta_{\mbox{bf}}$ is $\pi/4$ because for this value the result for $\sin^2 \theta_{13}$ is minimized: $\sin^2 \theta_{13} (\theta_{\mbox{bf}})=(2-\sqrt{3})/6 \approx 0.045$.
Notice that for $\theta_{\mbox{bf}} = \pi/4$ the Dirac phase $\delta$ is trivial and also one of the Majorana phases. The former is trivial because of a common $CP$
symmetry $Y=X_3$ of the matrices $m_l^\dagger m_l$ and $m_\nu^\dagger m_\nu$, as explained in subsection \ref{CPacc}. This $CP$ transformation fulfills eq.(\ref{XQcond}) for $Q=T$
and its form in the neutrino mass basis is 
\begin{equation}
Y_{\nu, \mbox{\tiny III} \normalsize} = U_\nu^\dagger X_3 U_\nu^\star= K^\star \left(
\begin{array}{ccc}
-i \sin 2 \theta &0 & i \cos 2 \theta\\
0 & -1 & 0\\
i \cos 2 \theta &0 & i \sin 2 \theta
\end{array}
\right) K^\star \; .
\end{equation}
 This matrix is for generic values of the parameter $\theta$ neither diagonal nor real. However,
if $\cos 2 \theta=0$ (as is true for $\theta=\pi/4$), $Y_{\nu,\mbox{\tiny III}\normalsize}$ becomes diagonal. 
Since its entries are not real (independently of $K$), $X_3$ is not a symmetry of the neutrino mass matrix itself but only of $m_\nu^\dagger m_\nu$ and consequently
Majorana phases are not trivial in general. According to eq. (\ref{Majofixed}), they can be directly read off from  $Y_{\nu, \mbox{\tiny III} \normalsize}$: $|\sin\alpha|=1$ and $\sin\beta=0$.
 
For the choice $G_e=Z_4$ we find only one case  which passes our selection criteria. We choose -as mentioned above- as representative $Z_4$ generating element $Q=S T U$ and 
take for $Z$ and $X$
 \begin{equation}
 Z=U \;\;\; \mbox{and} \;\;\; X=X_2 \; .
 \end{equation} 
 Notice that the same results would be obtained with the $CP$ symmetry $X_4$, since $X_4=Z X_2$.
The results for the mixing parameters and the best fit values $\theta_{\mbox{bf}}$ of the parameter $\theta$ can be found in table \ref{tab_Z4} and the mixing angles are displayed in figure 3.
 We notice the existence of two solutions, Case a and Case b, which arise from the freedom to exchange the second and the third rows of the PMNS matrix. 
The $\chi^2$ function of Case a has a global minimum
 with $\theta_{23} (\theta_{\mbox{bf}}) < \pi/4$, while that of Case b has a global minimum with $\theta_{23} (\theta_{\mbox{bf}}) > \pi/4$. These best fit points $\theta_{\mbox{bf}}$ are indicated as
 red dots in figure 3. In addition, we mark certain values of $\theta$ in order to guide the eye. 

 \begin{figure}[t!]
 \parbox{6.5in}{\includegraphics[scale=0.8]{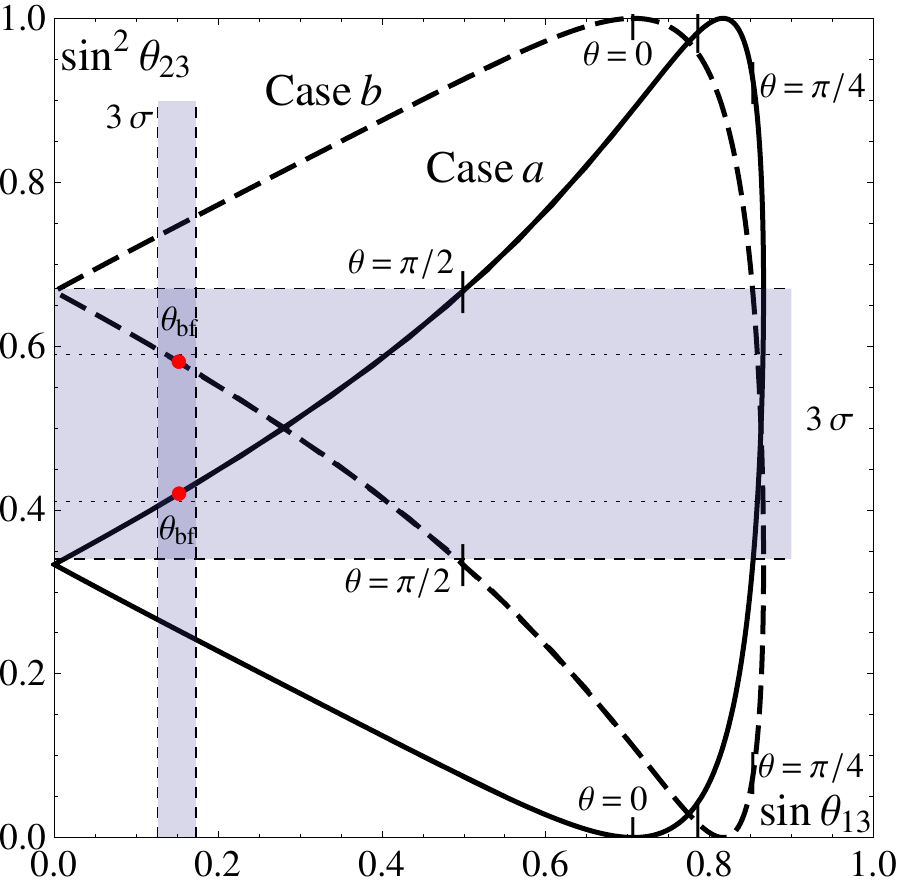}
 \includegraphics[scale=0.8]{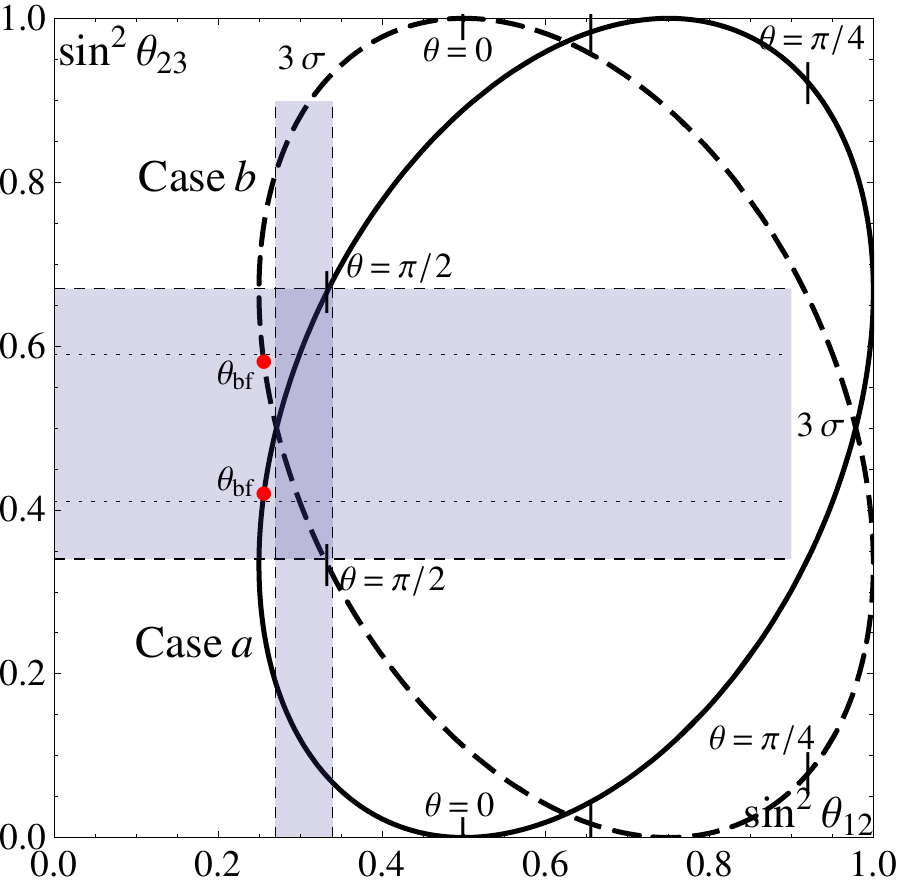}\\
 \includegraphics[scale=0.8]{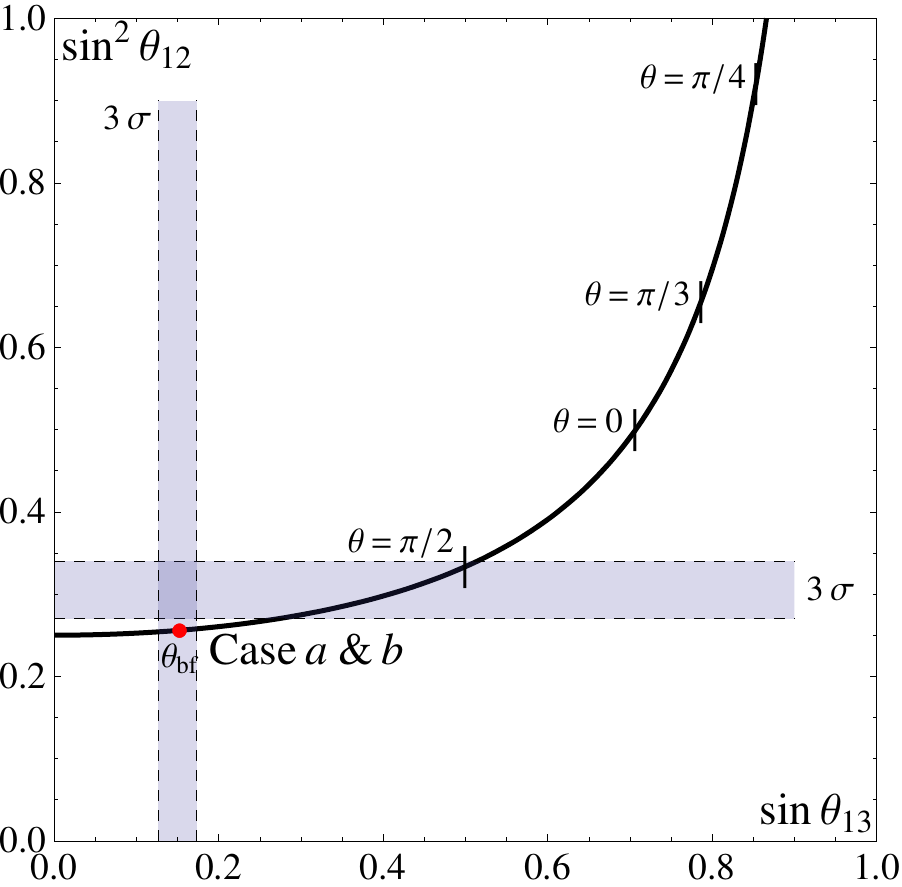} \parbox{3in}{\vspace{-2in}  {Figure 3: \small Results for the mixing angles $\sin \theta_{13}$, $\sin^2 \theta_{12}$
 and $\sin^2 \theta_{23}$ for Case a (straight line) and Case b (dashed line) for $G_e=Z_4$ or $G_e=Z_2 \times Z_2$. We mark the 
 value $\theta_{\mbox{bf}}$ of the parameter $\theta$ for which the $\chi^2$ functions have a global minimum with a red dot. For better guidance of the eye we also mark $\theta=0,\pi/4, \pi/3, \pi/2$ on the curves. 
 The shown $3 \, \sigma$ ranges for the mixing angles and the best fit values of the atmospheric mixing angle are taken from \cite{global_latest}.
 \normalsize}}}
 \end{figure}
 CP phases are trivial independently of the parameter $\theta$ which
 indicates the presence of an accidental $CP$ symmetry $Y$ in the charged lepton and neutrino sector. Indeed, the $CP$ symmetry $X=X_2$ of the neutrino sector is also a $CP$ symmetry of the charged
 lepton mass matrix $m^\dagger_l m_l$, because $Q=S T U$ and $X_2$ fulfill eq. (\ref{XQcond}).
 
 As one can check, the PMNS matrix has one column of the form $(1/2, 1/\sqrt{2}, 1/2)^T$ or $(1/2, 1/2, 1/\sqrt{2})^T$ which is in common (up to permutation) with the BM mixing pattern originating from the charged
 lepton sector (see eq.(\ref{UeQSTU}) for $G_e=Z_4$ and eq.(\ref{UeQ1Q2}) for $G_e=Z_2 \times Z_2$).
  Patterns with such a column have been recently mentioned in \cite{Hernandez_talk_BeNe}.
 
If a Klein group is preserved in the charged lepton sector and $G_\nu=Z_2 \times CP$, we find the same results as for $G_e$ being a $Z_4$ symmetry. One representative choice of the generators
of the Klein group in the charged lepton sector is
 \begin{equation}
 Q_1=T S T^2 S \;\;\; \mbox{and} \;\;\; Q_2=U T^2 
  \end{equation}
 and for $Z$ and $X$ of the neutrino sector we take
 \begin{equation}
 Z=U \;\;\; \mbox{and} \;\;\; X=X_1 \; .
 \end{equation}
All CP phases are trivial, because the imposed $CP$ symmetry of the neutrino sector $X=X_1$, the canonical $CP$ symmetry, is also present in the charged lepton
 sector. As one can check, the two generators $Q_1$ and $Q_2$ fulfill: $Q_i=Q_i^T$ with $i=1,2$ as required for $X=X_1$ by the constraint in eq.(\ref{XQcond}). 
The choice of $X_3$ as $CP$ symmetry in the neutrino sector leads to the same results as the choice of $X_1$, since $Z X_1=X_3$.
 
In appendix \ref{appB} we also discuss the mathematical structure of the groups arising from the combination of the group $S_4$ with one of the possible $CP$ transformations $X_i$.  
 
\subsection{Symmetry enhancement for particular values of $\theta$}
\label{symmenhance}

As has been mentioned at the beginning of subsection \ref{setup}, the maximal symmetry of a neutrino mass matrix $m_\nu$ for three generations of Majorana neutrinos is a Klein group, if
neutrino masses are unconstrained. In the present approach we assume in general only one of these $Z_2$ symmetries to be a subgroup of the flavour symmetry $G_f=S_4$.
(This is similar to the approach in \cite{GnuZ2_acc}.)
However, we notice that for the various cases presented particular values of the parameter $\theta$ (in general different for the different cases) exist for which the
second $Z_2$ symmetry of the neutrino mass matrix $m_\nu$ can be promoted to a subgroup of a finite flavour group which contains $S_4$. Thus, in these cases 
 the result of the mixing angles (and the CP phases) can be achieved through the symmetry breaking of this finite flavour group to a Klein group (and the symmetry $CP$) in the neutrino sector 
 instead to $Z_2 \times CP$. 

 In all cases displayed in table \ref{tab_c1_c5} the limit $\theta \rightarrow 0$ corresponds to TB mixing (and trivial CP phases). In this limit the second $Z_2$ symmetry of the 
neutrino sector turns out to be generated by $U$, belonging to $S_4$, so that the residual symmetry in the neutrino sector is $Z_2 \times Z_2 \times CP$ instead of only $Z_2 \times CP$.
The results of the mixing angles are thus the same as in the case without a $CP$ symmetry in the neutrino sector, see subsection \ref{A4S4Gfpatterns}.
In Case I also the limit $\theta \rightarrow \pi/4$ is noteworthy in which the mixing parameters take the values known from the democratic mixing matrix.
This can be understood, because the second $Z_2$ symmetry of the neutrino mass
matrix is now generated by $T S T^2 S$ so that again the neutrino sector is invariant under a residual symmetry $Z_2 \times Z_2 \times CP$. As mentioned in subsection \ref{groupS4}, $S$ and
$T$ generate the group $A_4$ and according to subsection \ref{A4S4Gfpatterns} the resulting mixing pattern is the democratic one. These special points are marked in figures
1 and 2 with a cyan dot and a green diamond, respectively. 

Interestingly, in Case II the PMNS matrix is of the form 
\begin{equation}
U_{PMNS}= \left( \begin{array}{ccc}
\sqrt{2/3} & 1/\sqrt{3} & 0\\
-1/\sqrt{6} & 1/\sqrt{3} & 1/\sqrt{2}\\
-1/\sqrt{6} & 1/\sqrt{3} & -1/\sqrt{2}
\end{array}
\right) \, R (\theta) \, K \; .
\end{equation}
This result should be compared with the findings of \cite{Delta96_384} in which it has been shown that the PMNS matrix takes this form (with a matrix $K$ with arbitrary phases, because Majorana phases are not 
determined in that approach) for $\theta=\pm \pi/12$ and $\theta=\pm \pi/24$, if the groups $\Delta (96)$ and $\Delta(384)$ are broken to a $Z_3$ symmetry in the
charged lepton sector and to a Klein group in the neutrino sector, respectively.

For Case III note that the results of mixing angles and the Dirac phase coincide for the best fit point $\theta_{\mbox{bf}}=\pi/4$
with those found in the case of a flavour group $\Delta (96)$ being broken to a Klein group in the neutrino sector and to a $Z_3$ symmetry in the charged lepton one \cite{Delta96_384}.
Indeed, we can check that for $\theta=\pi/4$ the second $Z_2$ symmetry under which the neutrino mass matrix $m_\nu$ is invariant can be generated by
\begin{equation}
\check{Z}=\left( \begin{array}{ccc}
 0 & 0 & i\\
 0 & 1 & 0\\
 -i & 0 & 0
\end{array}
\right) \; .
\end{equation}
Obviously, this matrix does not represent an element of the group $S_4$ in our real basis for $S$, $T$ and $U$. If we consider the group generated by $S$, $T$, $U$ and $\check{Z}$,
we find $\Delta (96)$ \cite{Delta6n2} using the computer program GAP \cite{gap}. Another particular value of the 
parameter $\theta$ is $\theta=0$, because in this case the element  $T S T^2 S$ is promoted to a  $Z_2$ generator under whose action the neutrino mass matrix
is invariant (independently of any $CP$ transformation). Both the residual
symmetry $G_e=Z_3$ in the charged lepton and $Z_2 \times Z_2$ in the neutrino sector are then generated through elements written in terms of $S$ and $T$ only so that the relevant flavour group is $A_4$
rather than $S_4$. As has been recapitulated in subsection \ref{A4S4Gfpatterns}, the flavour group $A_4$ broken in this way
leads to the democratic mixing pattern which coincides with the results of Case III for $\theta=0$.

For $G_e=Z_4$ one particular  limit is given by $\theta \rightarrow 0$ for which the (absolute values of the) PMNS matrix of Case a and Case b take(s) the form
 \be
 ||U_{PMNS, a}|| = \left( \begin{array}{ccc}
  1/2 & 1/2 & 1/\sqrt{2}\\
   1/\sqrt{2} & 1/\sqrt{2} & 0\\
  1/2 & 1/2 & 1/\sqrt{2}
 \end{array}
 \right)
  \; , \;\;
 ||U_{PMNS, b}|| = \left( \begin{array}{ccc}
  1/2 & 1/2 & 1/\sqrt{2}\\
    1/2 & 1/2 & 1/\sqrt{2}\\
   1/\sqrt{2} & 1/\sqrt{2} & 0
 \end{array}
 \right) \; ,
 \ee
 respectively.  As one recognizes, this is the BM mixing matrix up to permutations of rows and columns. From the viewpoint of group theory the value $\theta =0$ allows to promote $S$ to the  generator
 of one of the $Z_2$ symmetries of the matrix $m_\nu$ and thus the Klein group containing $S$ and $U$ is conserved in the neutrino sector. As repeated in  subsection \ref{A4S4Gfpatterns}, 
the choice $G_f=S_4$, $G_e=Z_4$ and $G_\nu$ being a Klein group leads to BM mixing.
If $G_e$ is also a Klein group, very similar statements hold: namely, $\theta=0$ promotes $S$ to the generator of a $Z_2$ symmetry of the matrix $m_\nu$ and the breaking of $G_f=S_4$ to $G_{e,\nu}$
being (non-normal) Klein groups with $G_e \neq G_\nu$ leads to BM mixing.

\subsection{Comments on $G_f=A_4$}
\label{commentA4}
   
 We can easily deduce results for $G_f=A_4$ from those obtained for $G_f=S_4$. 
 As mentioned in subsection \ref{groupS4} the generators $S$ and $T$ alone give rise to $A_4$. Thus, the only thing to do is to consider the above results
 which can be obtained with generators $Q_i$ and $Z$ made up of $S$ and $T$.
 This constrains us to take $G_e$ as $Z_3$ symmetry. Furthermore, only Case I, II and III are admissible, if $Z$ is required to be a product of $S$ and $T$. From the viewpoint
 of $A_4$ Case I and Case III are qualitatively different from Case II, since in the latter case $X$ is not proportional to an element of $A_4$, while it is for Case I and Case III. 

\section{Conclusions}
\label{concl}

The recent measurement of the reactor mixing angle $\theta_{13}$ has ruled out several of the prominent discrete flavour symmetries leading to 
  vanishing or very small $\theta_{13}$. This fact has led to several new ideas. 
For instance, we can still adopt a discrete flavour symmetry giving rise to TB mixing in first approximation.
However, in such a case, rather large corrections are needed in order to achieve the measured value of the reactor mixing angle.
At the same time, predictability is in general reduced, since such corrections tend to affect all mixing angles.
To retain predictability and to reproduce already in the lowest order approximation realistic mixing patterns
it is necessary to consider groups with larger order and several candidates have been found. Another possibility consists in reducing the residual symmetries
in the neutrino and/or charged lepton sector. In this case the symmetries do not completely determine all three mixing angles and it is possible to accommodate 
a non-vanishing value of $\theta_{13}$. However, the latter is then rather fitted than predicted.

In this paper we have analyzed a further possibility. We have considered a framework that is as predictive as possible, without necessarily requiring 
large discrete groups as starting point. Following earlier ideas found in the literature, we have included $CP$ in the set of symmetries of the theory
and have analyzed which role $CP$ could play in the symmetry breaking pattern. 
To define our framework we have first derived conditions that $CP$ and the flavour symmetries have to obey in order to provide a consistent set of transformations.
Given a flavour group, these conditions constrain the 
possible choice of $CP$ transformations. We have then discussed, in a model-independent way, a possible symmetry breaking pattern and its consequences for the prediction of the mixing parameters. 
Assuming three generations of Majorana neutrinos we have studied the case in which a flavour symmetry $G_f$ and a generalized $CP$ transformation are combined 
and are broken to $G_\nu=Z_2 \times CP$ in the neutrino sector, while $G_e \subset G_f$ remains preserved in the charged lepton one. 
Our main result is that all mixing angles and all CP phases, Dirac and Majorana, are determined in terms of a single real parameter $\theta$, 
which can take values between $0$ and $\pi$. 
This has to be compared with the approach in which $CP$ is absent, and thus Majorana phases cannot be constrained. 

In order to show concrete examples and find new interesting mixing patterns, we have performed
a comprehensive study for the group $G_f=S_4$. We have found several $CP$ transformations compatible with all our requirements, which lead to different results
for mixing angles and CP phases. Out of all possibilities only five give rise to phenomenologically interesting patterns, i.e. patterns for which the mixing angles are reasonably close to the best fit values for a certain value of the parameter $\theta$. 
In particular, a non-vanishing value of the mixing angle $\theta_{13}$ can be easily accommodated and all mixing parameters are strongly correlated.
Among the interesting cases we find two cases, Case I and Case IV in our classification, in which the neutrino mass matrix $m_\nu$ is invariant under $\mu\tau$ reflection symmetry  in the charged lepton mass basis. This particular generalized $CP$ transformation has already been studied in the literature. 
Surprisingly, several cases lead to trivial CP phases. For such cases we have figured out the presence of accidental $CP$ symmetries, not required a priori
by the assumed symmetry breaking pattern. In order to illustrate a case in which all CP phases are non-trivial functions of the parameter $\theta$, we have also discussed another case (Case III), which 
however cannot accommodate the best fit values of the mixing angles so well. Furthermore, we have briefly mentioned the cases which can also be obtained, if $G_f$ is $A_4$ instead of $S_4$.

Throughout this paper we have discussed the model-independent implications of our proposal. Physical results only depend on the assumed symmetry breaking pattern
and are independent of the details of a specific implementation, such as the particle content of the flavour symmetry breaking sector or the possible additional symmetries
of the theory. Nevertheless we think that it would be interesting to implement one of the presented cases in an actual model, for example in a supersymmetric context in which the breaking of the symmetry group to $G_e$ and $G_\nu$
is spontaneous due to non-vanishing vacuum expectation values of some flavons. In such a model for Case I or Case IV the prediction of a maximal Dirac phase $\delta$ for generic $\theta$ would not
depend on the parameters of the theory and thus could represent a concrete realization of geometrical CP violation.
Apart from the usual challenge that corrections can spoil some of the leading order predictions, models in which the presented idea is implemented 
have to face another challenge, namely the prediction of the parameter $\theta$ which depends on the entries of the neutrino mass matrix, and whose size is important for accommodating the
best fit values of the mixing angles well.

\section*{Acknowledgements}

FF and CH have been partly supported by the European Programme "Unification in the LHC Era", contract PITN-GA-2009-237920 (UNILHC). CH is also supported by the ERC Advanced Grant no. 267985,
"Electroweak Symmetry Breaking, Flavour and Dark Matter: One Solution for Three Mysteries" (DaMeSyFla).

\appendix

\section{Notation and proof of vanishing CP phases in presence of a generalized $CP$ transformation}
\label{appA}

In this appendix we first fix our notation and conventions for the mixing parameters and then prove that the presence of a generalized $CP$ transformation implies the vanishing of Dirac as well as Majorana phases.

\subsection{Notation}
\label{appA1}

As parametrization of the PMNS matrix we use  
\be
U_{PMNS} = \tilde{U} ~{\rm diag}(1, e^{i \alpha/2}, e^{i (\beta/2 + \delta)})~~~, 
\ee
with $\tilde{U}$ being of the form of the Cabibbo-Kobayashi-Maskawa (CKM) matrix $V_{CKM}$ \cite{pdg}
\be
\tilde{U} =
\begin{pmatrix}
c_{12} c_{13} & s_{12} c_{13} & s_{13} e^{- i \delta} \\
-s_{12} c_{23} - c_{12} s_{23} s_{13} e^{i \delta} & c_{12} c_{23} - s_{12} s_{23} s_{13} e^{i \delta} & s_{23} c_{13} \\
s_{12} s_{23} - c_{12} c_{23} s_{13} e^{i \delta} & -c_{12} s_{23} - s_{12} c_{23} s_{13} e^{i \delta} & c_{23} c_{13}
\end{pmatrix}
\ee 
and $s_{ij}=\sin\theta_{ij}$ and $c_{ij}=\cos\theta_{ij}$.  The mixing angles $\theta_{ij}$ range from $0$ to $\pi/2$, while the Majorana phases 
$\alpha, \beta$ as well as the Dirac phase $\delta$ take values between $0$ and $2 \pi$. The Jarlskog invariant 
$J_{CP}$ reads \cite{jcp}
\begin{eqnarray} \nonumber
J_{CP} &=&  {\rm Im} \left[ U_{PMNS,11} U_{PMNS,13}^* U_{PMNS,31}^* U_{PMNS,33}  \right] 
 \\ \label{JCP}
 &=& \frac 18 \sin 2 \theta_{12} \sin 2 \theta_{23} \sin 2 \theta_{13} \cos \theta_{13} \sin \delta \, .
\end{eqnarray}
Similar invariants, called $I_1$ and $I_2$, can be defined which depend on the Majorana phases $\alpha$ and $\beta$ 
\cite{Jenkins_Manohar_invariants} (see also \cite{rephasing_invariants_original,Majorana_invariants_other,Jenkins_Manohar_Hilbert}; see \cite{invariant_matrix_elements} for invariants
in terms of neutrino mass matrix elements)
\begin{eqnarray}
&&I_1 = {\rm Im} [U_{PMNS,12}^2 (U_{PMNS,11}^*)^2] = s^2_{12} c^2 _{12} c^4_{13} \sin \alpha~~~,
\\
&&I_2 =  {\rm Im} [U_{PMNS,13}^2 (U_{PMNS,11}^*)^2] = s^2 _{13} c^2 _{12} c^2_{13} \sin \beta~~~.
\end{eqnarray}
Notice that the Dirac phase has a physical meaning only if all mixing angles are different from $0$ 
and $\pi/2$, as indicated by the data. Analogously, the vanishing of the invariants $I_{1,2}$ only implies $\sin \alpha=0$, $\sin \beta=0$, if  solutions with $\sin2 \theta_{12} =0$, $\cos \theta_{13}= 0$ 
or $\sin 2 \theta_{13}=0$, $\cos \theta_{12}=0$ are discarded. Furthermore, notice that one of the Majorana phases becomes unphysical, if the lightest neutrino mass vanishes.

\subsection{Proof of vanishing CP phases in presence of a generalized $CP$ transformation}

The presence of a generalized $CP$ transformation $X$ under which charged lepton as well as neutrino mass matrices are invariant leads to the vanishing of all three
CP invariants $J_{CP}$, $I_{1}$ and $I_{2}$ and thus to trivial Dirac and Majorana phases.  In the following, we focus on the case of Majorana neutrinos, but also briefly comment on Dirac neutrinos.
All these results are known in the literature, see e.g. \cite{Branco_review}.

In order to study the consequences of $CP$ invariance on the lepton mixing parameters, we diagonalize the mass matrices $m_l$ and $m_\nu$ by unitary matrices $U_e$ and $U_\nu$, respectively,
\be
 m_l^\dagger m_l = U_e  (m_l^\dagger m_l)^{diag} U_e^\dagger~~~,
\label{ldiag}
\qquad  \qquad
 m_\nu = U_\nu^* m_\nu^{diag} U_\nu^\dagger~~~,
 \ee
 so that the invariance conditions in eq. (\ref{invcond}) in subsection \ref{CPgen} become
 \be
 \label{condK}
 X_e^* (m_l^\dagger m_l)^{diag}  X_e = (m_l^\dagger m_l)^{diag}~~~, \qquad  \qquad
 X_\nu m_\nu^{diag} X_\nu = m_\nu^{diag}~~~,  
 \ee
 with
 \be 
 \label{Xcond}
 X_e = U_e^\dagger X U_e^*~~~, \qquad  \qquad X_\nu = U_\nu^\dagger X U_\nu^*~~~.
 \ee
The matrices $X_e$ and $X_\nu$ represent the $CP$ transformation $X$ in the lepton mass basis. For non-degenerate charged lepton and neutrino masses one can show that the general solutions of eq.~(\ref{condK})
for $X_e$ and $X_\nu$ are diagonal matrices
\be
\label{conddiag}
X_e = \begin{pmatrix} e^{i x_1} & 0 & 0 \\ 0 & e^{i x_2} & 0 \\ 0 & 0 & e^{i x_3} \end{pmatrix}~~~, \qquad \qquad X_\nu = \begin{pmatrix} s_1 & 0 & 0 \\ 0 & s_2 & 0 \\ 0 & 0 & s_3 \end{pmatrix}
\ee
with
\be
\label{condXphases}
x_i \in [ 0, 2 \pi ) ~~~, \qquad\qquad s_i = \pm 1 ~~~.
\ee
This can be stated differently: the lepton sector is invariant under a generalized $CP$ transformation described by $X$, if one can find matrices $X_e$ and $X_\nu$ of the form as in eqs.(\ref{conddiag}, \ref{condXphases}), 
such that
\be
\label{supercond}
U_\nu X_\nu U_\nu^T = U_e X_e U_e^T~~~.
\ee 
Provided that such $X_e$ and $X_\nu$ exist, we can study  the consequences for the PMNS matrix which is given by
\be
\label{UPMNS}
U_{PMNS} = U_e^\dagger U_{\nu} \; . 
\ee
From eqs.(\ref{supercond},\ref{UPMNS}) we see that
\be
\label{XUPMNScond}
X_e^* U_{PMNS} X_\nu = U_{PMNS}^\star
\ee
holds. Plugging in eqs.(\ref{conddiag},\ref{condXphases}), we find
\cite{CPconv_cond}
\be
\label{PMNSCPcon}
U_{PMNS, ij} \, e^{-i x_i} s_j =  U^*_{PMNS, ij} \; .
\ee 
The Jarlskog invariant $J_{CP}$ in eq.(\ref{JCP}) and $\sin\delta$ then vanish. Also $I_1$ and $I_2$ vanish, if the PMNS matrix fulfills eq.(\ref{PMNSCPcon}). 
Thus, also the Majorana phases $\alpha$ and $\beta$ are trivial in this case. These conclusions hold unless the mixing angles take special values, as commented at the end of appendix \ref{appA1}.

For Dirac neutrinos, $m_\nu$ and $X_\nu$ are subject to the same constraints as the matrices $m_l$ and $X_e$, see eq.(\ref{Diracnu}), and thus $m_\nu$
and $X_\nu$ have to satisfy (only)
\be
\label{CPDirac1}
 X_\nu^* (m_\nu^\dagger m_\nu)^{diag}  X_\nu = (m_\nu^\dagger m_\nu)^{diag}
\ee
instead of eqs.(\ref{invcond},\ref{condK}). As a consequence, the most general form of $X_\nu$ fulfilling eq.(\ref{CPDirac1}) for a non-degenerate neutrino mass spectrum
is the same as for $X_e$ in eqs.(\ref{conddiag},\ref{condXphases}) (with $x_i$ called $x^\nu_i$). If this is realized, eq.(\ref{XUPMNScond}) implies
\be
\label{CPDirac2}
U_{PMNS, ij} \, e^{-i (x_i-x^\nu_j)} =  U^*_{PMNS, ij} 
\ee
and thus a vanishing Jarlskog invariant (and trivial Dirac phase).

\section{Mathematical structure of $G_f$ and $CP$}
\label{appB}

In this appendix, we comment on the mathematical structure of the group comprising $G_f$ and $CP$ and show that it is in general a semi-direct product. We assume
to be given a solution $X$ of the constraint in eq. (\ref{Xcond2}) and that ${\bf r}$ is a faithful representation\footnote{In this way, every element of the abstract group $G_f$ is represented by a different
representation matrix.} of $G_f$ (which does not need to be discrete and/or finite in the following). In order to analyze which group arises 
from combining the transformations of $G_f$ and those of $H_{CP}$, the parity group generated by $CP$,
we enlarge the field space from $\varphi$ to $\Phi=(\varphi,\varphi^*)^T$. In this space the actions of $G_f$ and $CP$ are given by:
\be
\Phi'={\cal A} \Phi~~~~~~~~~~\Phi'={\cal X} \Phi~~~,
\ee
with
\be
\label{calAX}
{\cal A}=
\begin{pmatrix}
A&0\\
0&A^*
\end{pmatrix}~~~~~~~~~~{\cal X}=
\begin{pmatrix}
0&X\\
X^*&0
\end{pmatrix}~~~,
\ee 
respectively. The unitary matrices ${\cal A}$ generate a group which is isomorphic to $G_f$. The matrix ${\cal X}$ satisfies 
\be
{\cal X}^2=\mathbb{1}~~~,
\label{XCOND1}
\ee
since $X$ is unitary and symmetric, and thus generates a group $\{{\cal X},\mathbb{1}\}$ isomorphic to $H_{CP}$.
We do not distinguish between two isomorphic groups and we use the same notation for both.
The consistency condition in eq. (\ref{Xcond2}) reads
\be
{\cal X}^{-1} {\cal A}{\cal X}={\cal A}'~~~.
\label{XCOND2}
\ee
The group $G_{CP}$ we are looking for is that
generated by the unitary matrices ${\cal A}$ and ${\cal X}$. We observe that the closure of this set of transformations is guaranteed by eqs. (\ref{XCOND1},\ref{XCOND2}).
Indeed by using these equations we can show that any element of $G_{CP}$ can be cast into the form ${\cal A}{\cal H}$ with ${\cal A}$ belonging to $G_f$
and ${\cal H}$ belonging to the group $H_{CP}$.  Such a decomposition is unique, namely ${\cal A}{\cal H}={\cal A'}{\cal H'}$ implies ${\cal A'}={\cal A}$ and ${\cal H'}={\cal H}$. This is equivalent to the 
statement that the intersection of the two groups $G_f$ and $H_{CP}$ is only the neutral element. The multiplication rule
of two elements of $G_{CP}$ follows from eqs.  (\ref{XCOND1},\ref{XCOND2}) and is given by 
\be
{\cal A}{\cal H}~{\cal A'}{\cal H'}={\cal A}{\cal A''}~{\cal H}{\cal H'}~~~~~~~~~~~~~~~{\cal A''}={\cal H}{\cal A'}{\cal H}~~~.
\ee

Therefore the group $G_{CP}$ is isomorphic to the semi-direct product of $G_f$ and $H_{CP}$, $G_{CP}=G_f\rtimes H_{CP}$ and $H_{CP}$ acts on $G_f$. $G_f$ is a normal subgroup of $G_{CP}$, while $H_{CP}$ is in general only a subgroup. 
If $G_f$ is a finite group, the group $G_{CP}$ has twice as many elements as $G_f$. The semi-direct product reduces to a direct one if and only if  ${\cal X}$ commutes with all elements of $G_f$, 
that is when the condition in eq. (\ref{XCOND2}) is satisfied with ${\cal A'}={\cal A}$ for all elements ${\cal A}$ of $G_f$. (This is equivalent to the case in which the condition in eq. (\ref{Xcond2}) is satisfied with $A'=A$ for all $A$.)
Then the subgroup $H_{CP}$ is also normal. 

We note that under a change of basis
\be
\tilde{\Phi}={\cal O}^\dagger\Phi~~~~~~~~~~{\cal O}=
\begin{pmatrix}
\Omega &0\\
0&\Omega^*
\end{pmatrix}
\ee
with ${\cal O}$ unitary, all elements of $G_{CP}$ have the same transformation properties, as deduced from eqs.(\ref{change1},\ref{change2},\ref{calAX}),
\be
\tilde{\cal X}={\cal O}^\dagger{\cal X}{\cal O}~~~,~~~\tilde{\cal A}={\cal O}^\dagger{\cal A}{\cal O}~~~.
\ee 

\vspace{0.15in}

In the particular case $G_f=S_4$, for which we discuss in detail the possible choices of $X$ as well as the phenomenological results for lepton mixing angles and CP phases in section \ref{sec3},
we find the following:  among the various $X_i$, see eqs.(\ref{admissX_1}, \ref{admissX_2}), only  the canonical $CP$ transformation $X=X_1$ leads to a direct product $S_4 \times CP$, because
only in this case $A X=X A$ holds for all generators $A$ (and thus all elements) of $S_4$ (we have used the fact that all representation matrices are real). In all the other
 cases $A X= X A$ is only satisfied for one or two of the three generators $S$, $T$ and $U$. In particular, for $X=X_2 \propto S$, $X=X_3 \propto U$ and $X=X_4 \propto S U$ the relation $A X=X A$
 obviously holds for $A=S$ and $A=U$, whereas for $A=T$ we find $X^{-1} T X =S T S$, $T^2$ and $T S T$, respectively. For the two $CP$ transformations
 $X=X_5$ and $X=X_6$ $A X=X A$ is only fulfilled for the generator $A=S$, as required in order to preserve the direct product of
 the $Z_2$ symmetry generated by $S$ and of the $CP$ transformation $X$ in the neutrino sector, see eq.(\ref{xz}). For $A=U$ we find instead $X^{-1} A X= S U$ in both cases and  
 $A=T$ gives rise to $X^{-1} A X=T S$ and $S T$, respectively.


\end{document}